\documentclass[journal]{IEEEtran}
\usepackage{amsfonts}
\usepackage{amsmath}
\usepackage{amssymb}
\usepackage{cite}
\usepackage{bm}
\usepackage{graphicx}
\usepackage[dvipsnames]{xcolor}
\usepackage{booktabs}
\usepackage{subfigure}
\usepackage{multirow}
\usepackage{color, colortbl}
\usepackage{url}
\usepackage{multirow}
\usepackage{makecell}
\usepackage[colorlinks, linkcolor=red, anchorcolor=blue, citecolor=green]{hyperref}
\usepackage{mathtools}
\usepackage[ruled,vlined]{algorithm2e}
\usepackage{esvect}

\newcommand{\RR}[2]{\textcolor[rgb]{0,0,0}
{#2}}

\definecolor{LightCyan}{rgb}{0.49, 0.78, 0.91}

\ifCLASSINFOpdf
\else
\fi

\begin{document}

\twocolumn[
  \begin{@twocolumnfalse}
    This work has been submitted to the IEEE for possible publication. Copyright may be transferred without notice, after which this version may no longer be accessible.
  \end{@twocolumnfalse}
]

\newpage

\title{Deep Reinforcement Learning for Band Selection in Hyperspectral Image Classification}
\author{Lichao~Mou,
        Sudipan~Saha,~\IEEEmembership{Member,~IEEE,}
        ~Yuansheng~Hua,
        ~Francesca~Bovolo,~\IEEEmembership{Senior Member,~IEEE,}
        ~Lorenzo~Bruzzone,~\IEEEmembership{Fellow,~IEEE,} and
        ~Xiao~Xiang~Zhu,~\IEEEmembership{Fellow,~IEEE}

\thanks{This work is jointly supported by the European Research Council (ERC) under the European Union's Horizon 2020 research and innovation programme (grant agreement No. [ERC-2016-StG-714087], Acronym: \textit{So2Sat}), by the Helmholtz Association
through the Framework of Helmholtz AI - Local Unit ``Munich Unit @Aeronautics, Space and Transport (MASTr)'' and Helmholtz Excellent Professorship ``Data Science in Earth Observation - Big Data Fusion for Urban Research'' and by the German Federal Ministry of Education and Research (BMBF) in the framework of the international future AI lab ``AI4EO -- Artificial Intelligence for Earth Observation: Reasoning, Uncertainties, Ethics and Beyond'' (Grant number: 01DD20001).}
%\thanks{L. Mou is with the Remote Sensing Technology Institute (IMF), German Aerospace Center (DLR), Germany (e-mail: lichao.mou@dlr.de).}
\thanks{L. Mou, Y. Hua, and X. X. Zhu are with the Remote Sensing Technology Institute (IMF), German Aerospace Center (DLR), Germany and with the Data Science in Earth Observation (SiPEO, former: Signal Processing in Earth Observation), Technical University of Munich (TUM), Germany (e-mails: lichao.mou@dlr.de; yuansheng.hua@dlr.de; xiaoxiang.zhu@dlr.de).}
\thanks{S. Saha is with the Data Science in Earth Observation (SiPEO, former: Signal Processing in Earth Observation), Technical University of Munich (TUM), Germany (e-mail: sudipan.saha@tum.de).}
\thanks{F. Bovolo is with the Fondazione Bruno Kessler, Italy (email: bovolo@fbk.eu).}
\thanks{L. Bruzzone is with the Department of Information Engineering and Computer Science, University of Trento, Italy (email: lorenzo.bruzzone@ing.unitn.it).}
       }

\maketitle

\begin{abstract}
\textcolor[rgb]{0,0,1}{This is the preprint version. To read the final version, please go to IEEE Transactions on Geoscience and Remote Sensing.} Band selection refers to the process of choosing the most relevant bands in a hyperspectral image. By selecting a limited number of optimal bands, we aim at speeding up model training, improving accuracy, or both. It reduces redundancy among spectral bands while trying to preserve the original information of the image. By now many efforts have been made to develop unsupervised band selection approaches, of which the majority are heuristic algorithms devised by trial and error. In this paper, we are interested in training an intelligent agent that, given a hyperspectral image, is capable of automatically learning policy to select an optimal band subset without any hand-engineered reasoning. To this end, we frame the problem of unsupervised band selection as a Markov decision process, propose an effective method to parameterize it, and finally solve the problem by deep reinforcement learning. Once the agent is trained, it learns a band-selection policy that guides the agent to sequentially select bands by fully exploiting the hyperspectral image and previously picked bands. Furthermore, we propose two different reward schemes for the environment simulation of deep reinforcement learning and compare them in experiments. This, to the best of our knowledge, is the first study that explores a deep reinforcement learning model for hyperspectral image analysis, thus opening a new door for future research and showcasing the great potential of deep reinforcement learning in remote sensing applications. Extensive experiments are carried out on four hyperspectral data sets, and experimental results demonstrate the effectiveness of the proposed method. The code is publicly available\footnote{\url{https://github.com/lcmou/DRL4BS}}.
\end{abstract}


\IEEEpeerreviewmaketitle

\section{Introduction}
\label{sec:intro}
\IEEEPARstart{I}{n} remote sensing, spectral sensors are widely used for Earth observation tasks, like land cover classification~\cite{camps2013advances,gu2017multiple,he2017recent,audebert2019deep,li2019deep,camps2005kernel,shi2013semisupervised,li2013generalized,mou2017deep,li2016hyperspectral,mou2017unsupervised,lgrs/HautPPPL19,zhong2018spectral,PaolettiHFPPP19,ssun}, anomaly detection~\cite{tgrs/LiD15,tgrs/ZhangDZW16,tcyb/YuanMW16,tgrs/KangZLLLB17,huang2020subpixel}, and change detection~\cite{tgrs/BruzzoneS97,DBLP:journals/tgrs/BovoloB07,tgrs/BovoloMB12,/tgrs/WuDZ14,bovolo2015time,tip/ZanettiBB15,remotesensing/LyuLM16,wang2018getnet,tgrs/MouBZ19,tgrs/DuRWZ19,tgrs/ZhaoMCBE20,tgrs/SahaBB19,saha20}. A hyperspectral image often comprises hundreds of spectral bands within and beyond the visible spectrum. Such an image can be deemed as a hyper-cube, providing rich spectral information that helps to identify various land covers. Hyperdimensionality also raises some issues, e.g., a high level of redundancy among spectral bands, high computational overheads, and large storage requirements. Therefore, it is beneficial to reduce data redundancy.
\par
In the literature, two kinds of methodologies, namely feature extraction~\cite{icadr06,BandosBC09} and band selection~\cite{ChangDSA99,DuQWRS03,XieJARS17,mev,ZhangLDZ18,UsoPSG07,JiaTZL16,WangZL18,WangLL19,Roy2020,yin2010optimal,chang2014hyperspectral,tnn/WangLY16,sun2015band,sun2020fast,tgrs/PatraMB15,WSun19,tgrs/0001BZTG18,ZhangGJZ18,JFeng2020}, are commonly used to reduce redundancy in hyperspectral images. The former transforms original hyperspectral data into a lower dimension via a linear or nonlinear mapping. For example,~\cite{icadr06} makes use of independent component analysis (ICA) to extract features from a hyperspectral image in an unsupervised way. In~\cite{BandosBC09}, the authors investigate a supervised feature extraction approach based on linear discriminant analysis (LDA). Moreover, several works put effort into using manifold learning algorithms, e.g., Laplacian eigenmaps (LE)~\cite{le}, locally linear embedding (LLE)~\cite{lle}, and isometric feature mapping (Isomap)~\cite{isomap}, to learn low-dimensional features by taking advantage of the underlying geometric structure of hyperspectral data. On the other hand, band selection refers to the process of choosing a cluster of informative spectral bands and discarding ones that are often not discriminative enough for the considered problem. Unlike feature extraction, band selection can keep the physical meaning of original hyperspectral images and be better interpreted for certain tasks~\cite{WSun19}. Hence in this paper, we are interested in hyperspectral band selection. \RR{}{Band selection is applicable to tasks as diverse as hyperspectral image classification, change detection, and anomaly detection. In this work, we use classification tasks to validate the effectiveness of selected bands.}
\par
From the perspective of the availability and use of labeled data, band selection methods are grouped into the following three categories: unsupervised, semi-supervised, and supervised. Semi-supervised and supervised models exploit labeled samples to learn a band selection strategy. Such labeled data, however, are not often available in practical remote sensing applications. Hence unsupervised band selection is more desirable in the community. In this direction, the existing methods can be approximately sorted into the following categories:
\begin{itemize}
  \item Ranking-based methods. These methods aim at seeking an effective criterion to measure the significance of each spectral band and prioritize all bands. Afterwards, top-ranked bands are selected. Some representative ranking-based band selection methods are~\cite{ChangDSA99,DuQWRS03,XieJARS17}.
  \item Searching-based methods. The searching-based band selection approaches usually have two components: an objective function and a sequential search algorithm. The former is a criterion that the latter seeks to minimize over all feasible band subsets by adding or removing bands from a candidate set. The searching-based methods have two variants: sequential forward selection and sequential backward selection. \cite{mev,ZhangLDZ18} are both representative works in this direction.
  \item Clustering-based methods. In these methods, all spectral bands are first grouped into several clusters via a clustering algorithm. Afterwards, the most representative band is selected from each cluster. Representative clustering-based band selection methods include~\cite{UsoPSG07,JiaTZL16,WangZL18,WangLL19}.
  \item Others. Some hybrid approaches, e.g., combining ranking and clustering~\cite{yin2010optimal,chang2014hyperspectral,tnn/WangLY16}, are proposed for band selection tasks. Furthermore, sparse learning, low rank representation, and deep learning also provide new insights~\cite{sun2015band,sun2020fast,Roy2020}.
\end{itemize}
\par
In essence, hyperspectral band selection can be treated as a combinatorial optimization problem. The aforementioned methods that use exact and heuristic algorithms have proven to be effective for such a task. However, these heuristic algorithms are devised based on domain knowledge from human experts by trial and error. Hence we are curious as to whether this heuristic design procedure for unsupervised band selection tasks can be automated using artificial intelligence techniques. If feasible, there would be much to be gained. Reinforcement learning systems are trained from their own experience, in principle allowing them to operate in tasks where human expertise is lacking and thus being suitable for discovering new band selection methods without any hand-engineered reasoning. Recently deep reinforcement learning, introducing deep learning into reinforcement learning, has demonstrated breakthrough achievements in various fields~\cite{MnihKSRVBGRFOPB15,CaicedoL15,SilverHMGSDSAPL16,sallab2017deep,shao2019survey}. In this paper, we propose a framework that can solve the problem of unsupervised band selection using deep reinforcement learning.

This work's novel contributions are in the following aspects:
\begin{itemize}
  \item We cast the problem of unsupervised hyperspectral band selection as a Markov decision process of an agent and then solve this problem with a deep reinforcement learning algorithm. To the best of our knowledge, this is the first study that makes use of deep reinforcement learning for the task of band selection.
  \item We propose an effective solution to parameterize the Markov decision process for optimal band selection. More specifically, for the agent, we devise the set of actions, the set of states, and an environment simulation tailored for this task.
  \item We present and discuss two instantiations of the reward scheme of the environment simulation, namely information entropy and correlation coefficient, for unsupervised hyperspectral band selection.
  \item We train a deep reinforcement learning model using Q-network to learn a band-selection policy whose effectiveness has been validated extensively with various data sets and classifiers.
\end{itemize}
\par
We organize the remainder of this paper as follows. Hyperspectral band selection is detailed in Section~\ref{sec:intro}. Section~\ref{sec:method} introduces the proposed model. Section~\ref{sec:exp} tests the proposed model and presents experimental results as well as the discussion. Lastly, the paper is concluded in Section~\ref{sec:conc}.

\begin{figure}[!t]
\centering
\includegraphics[width=\columnwidth]{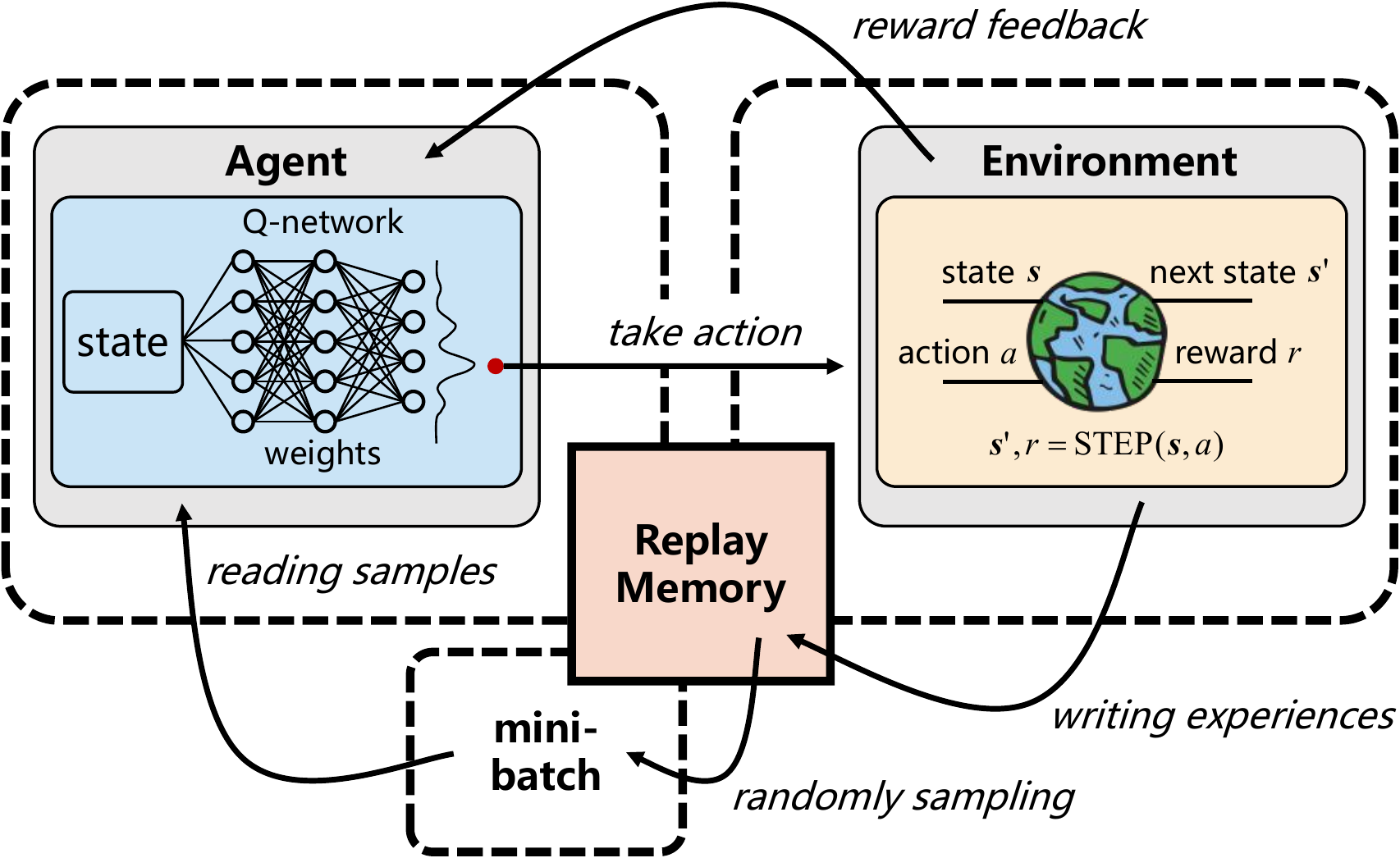}
\renewcommand{\figurename}{Fig.}
\caption{\label{fig:drl} An overview of the proposed deep reinforcement learning model for unsupervised hyperspectral band selection. In the training phase, an intelligent agent (Q-network) interacts with a tailored environment in order to learn a band-selection policy by trial and error. Specifically, the Q-network takes as input the state representation encoding selected bands and outputs a vector whose each component is a Q-value for each band. In the test phase, the agent selects bands according to the learned policy.}
\end{figure}

\section{Methodology}
\RR{}{Let us consider a hyperspectral image with $L$ bands. Our goal is to select $K$ optimal bands to reduce redundancy. The number of all possible combinations is $\tbinom{L}{K}$. Suppose that $L=200$ and $K=30$, the number is about $4\times10^{35}$.} In this work, we first formulate the task as a Markov decision process, as detailed in Section~\ref{subsec:MDP}. Afterwards, a deep reinforcement learning model is used to solve this problem (see Section~\ref{subsec:drl}). Section~\ref{subsec:imp} discusses implementation details.

\label{sec:method}
\subsection{Problem Formulation: Band Selection as a Markov Decision Process}
\label{subsec:MDP}
We view the task of hyperspectral band selection as a sequential forward search process, i.e., a sequential decision-making problem of an agent which interacts with a tailored environment (cf.~Fig.~\ref{fig:drl}). To be more specific, the agent needs to decide which spectral band it should pick at each time step so that it can find an optimal combination of $K$ bands in $K$ steps, and during this procedure, the agent explores the environment through actions and observes rewards and states. In this paper, we cast this problem as a Markov decision process that offers a formal framework for modeling the procedure of sequential decision-making when outcomes are partially uncertain.
\par
A 5-tuple $(\mathcal{A},\mathcal{S},P,R,\gamma)$ is often used to define a Markov decision process~\cite{Bellman1957}. Here $\mathcal{A}$ denotes the set of all actions, $\mathcal{S}$ is the set of all countable or uncountable states, $P:\mathcal{S}\times \mathcal{A}\rightarrow \mathcal{P}(\mathcal{S})$ represents Markov transition function, $R:\mathcal{S}\times \mathcal{A}\rightarrow \mathcal{P}(\mathbb{R})$ is the distribution of immediate rewards of state-action pairs, and $\gamma\in(0,1)$ denotes a discount factor. In specific, upon taking an action $a\in\mathcal{A}$ at a state $\bm{s}\in\mathcal{S}$, the probability distribution of the next state can be defined by $P(\cdot|\bm{s},a)$, and $R(\cdot|\bm{s},a)$ depicts the distribution of the immediate reward for the chosen action. In what follows, we detail how we parameterize the Markov decision process for our case.
\par
\textbf{Action.} The action of the agent in our case is to choose a spectral band from the hyperspectral image at each time step. The complete set of all actions $\mathcal{C}$ is identical to the set of bands, i.e., $\mathcal{C}=\{1,2,\cdots,L\}$. Let $\mathcal{B}$ be a set consisting of actions that have been taken before. Then the actual set of actions for the current time step is $\mathcal{A}=\mathcal{C}\setminus\mathcal{B}$. During the training phase, an action $a$, $a\in\mathcal{A}$, is taken by the agent and subsequently sent to the environment, and the latter receives the action, evaluates it, and gives the agent a positive or negative reward. In the test phase, the agent acts according to a learned policy to sequentially select bands.
\par
\textbf{State.} The state $\bm{s}$ in our case is represented as the action history of the agent and is denoted as a $L$-dimensional vector with multi-hot encoding that records which actions have been taken (i.e., which spectral bands have been chosen) in the past. For example, $s_i=1$ means that the $i$-th band has been picked in previous time steps, while $s_i=0$ represents that it is still selectable. Taking the action history as the state implies dependencies among spectral bands, which helps to select the next band. Note that there exists a one-to-one correspondence between $\bm{s}$ and $\mathcal{B}$.
\par
\textbf{Transition.} The transition function $P$ deems the next state as a possible outcome of taking an action at a state. In this work, the transition function is deterministic, which means that the next state is specified for each state-action pair. Specifically, $P$ updates the state by changing the action history as follows:
\begin{equation}
\label{eq:5}
\mathcal{B}^{\prime}=\mathcal{B}\cup\{a\} \,,
\end{equation}
where $\mathcal{B}^{\prime}$ represents the set of selected bands associated with the next state $\bm{s}^{\prime}$.
\par
\textbf{Reward.} The reward function $R$ should be in proportion to the advancement that the agent makes after picking a specific band. In this work, we discuss two ways to instantiate our reward scheme and measure the improvement from one state to another in our setup. They are detailed as follows.
\begin{itemize}
  \item Information entropy: The information entropy is capable of measuring the information amount of a random variable quantitatively. Hence we make use of it to evaluate the richness of spectral information of bands. More specifically, denote $\bm{x}_i\in\mathbb{R}^N$ as the $i$-th band vector, we calculate the mean information entropy of selected bands as follows:
      \begin{equation}
      \label{eq:1}
      \text{MIE}(\bm{s})=-\frac{1}{|\mathcal{B}|}\sum_{i\in\mathcal{B}}\sum_{n=1}^NP(x_i^n)\log_2P(x_i^n) \,,
      \end{equation}
      where $\mathcal{B}$ is associated with the state $\bm{s}$. When the agent takes an action $a$ and moves from state $\bm{s}$ to $\bm{s}^{\prime}$, the reward $R(\bm{s},\bm{s}^{\prime})$ can be calculated as follows:
      \begin{equation}
      \label{eq:2}
      R(\bm{s},\bm{s}')=\text{MIE}(\bm{s}^{\prime})-\text{MIE}(\bm{s}) \,.
      \end{equation}
  \item Correlation coefficient: The correlation coefficient measures how strong the relationship between two variables is. Here, we use it to estimate intra-band correlations among selected bands. There are several types of correlation coefficients, and we exploit a commonly used one, Pearson's correlation, a.k.a., Pearson's R, to calculate the mean correlation coefficient for $\bm{s}$ as follows:
      \begin{equation}
      \label{eq:3}
      \text{Corr}(\bm{s})=\frac{1}{|\mathcal{B}|^2}\sum_{i\in\mathcal{B}}\sum_{j\in\mathcal{B}}\frac{E[(\bm{x}_i-\mu_{\bm{x}_i})(\bm{x}_j-\mu_{\bm{x}_j})]}{\sigma_{\bm{x}_i}\sigma_{\bm{x}_j}} \,.
      \end{equation}
      Then the agent can be rewarded by the following formula:
      \begin{equation}
      \label{eq:4}
      R(\bm{s},\bm{s}') = \text{Corr}(\bm{s})-\text{Corr}(\bm{s}^{\prime}) \,.
      \end{equation}
\end{itemize}
Intuitively, Eq.~(\ref{eq:2}) and Eq.~(\ref{eq:4}) tell that the reward is positive if the quality of selected bands is improved from state $\bm{s}$ to state $\bm{s}^{\prime}$, and negative otherwise. Driven by this reward scheme, the agent pays a penalty for choosing a non-informative band and is rewarded to add a band that results in an increase in the informative content of the whole set of selected bands. We quantitatively compare the above two instantiations of the reward scheme in Section~\ref{subsec:exp-c}. The information entropy and correlation coefficient are two commonly used metrics to assess the quality of bands selected by an unsupervised band selection model~\cite{JiaTZL16,sun2020fast}, which is the reason why we consider them as the reward scheme. Furthermore, we believe that more alternatives are possible and may improve results in the future.

\subsection{Deep Reinforcement Learning for Band-selection Policy}
\label{subsec:drl}
In Section~\ref{subsec:MDP}, we discuss the parameterization of the Markov decision process for our task. By doing so, the band selection task is transformed into a sequential decision-making problem. Next, we show how we use deep reinforcement learning to learn a band-selection policy in this setup.
\par
The policy we seek is an action-value function, denoted by $Q(\bm{s},a)$\footnote{``Q'' in reinforcement learning is an abbreviation of the word ``Quality''.}, that specifies the action $a$ to be taken when the current state is $\bm{s}$. Based on this function, the agent chooses the action which is associated with the maximum reward value. That is to say, in our task, bands with high information entropy or low correlation are expected to be chosen. Q-learning~\cite{Watkins1989}, a classical reinforcement learning algorithm, is often employed to approximate $Q(\bm{s},a)$ by iteratively updating the action-selection policy using the Bellman equation:
\begin{equation}
\label{eq:bellman}
Q(\bm{s},a)=r+\gamma\max_{a^{\prime}}Q(\bm{s}^{\prime},a^{\prime}) \,,
\end{equation}
where $r$ denotes the immediate reward and the second term $Q(\bm{s}^{\prime},a^{\prime})$ is a future reward. In Q-learning, a lookup table, termed Q-table, serves as the Q-function for the agent to query the best action. However, this becomes impractical when action and state spaces are very large. To tackle such a problem, in this paper, we exploit a network named Q-network to approximate the action-value function.
\par
\textbf{Q-network architecture.} The Q-network takes as input the state representation introduced in Section~\ref{subsec:MDP} and outputs a vector whose each component is a Q-value for each action. A detailed description of the Q-network we use is as follows. The input consists of a $L$-dimensional vector. The first fully connected layer has 2$L$ units, followed by rectifier linear units (ReLUs)~\cite{KrizhevskySH12}. The second fully connected layer has the same structure as the first layer, again followed by ReLUs. Finally, the last layer, a linear fully connected layer with $L$ units, follows. \RR{}{The structure of the Q-network is outlined in Table~\ref{tab:qnet_archi}.}
\par

\begin{algorithm}[t]
\label{algo1}
\SetAlgoLined
 randomly initialize Q-network weights $\bm{w}$\;
 initialize replay memory $\mathcal{M}$\;
 initialize the complete set of all actions $\mathcal{C}$\;
 \While{not converged}
 {
 initialize state: $\bm{s}=\vec{0}$\;
 empty the set of chosen bands: $\mathcal{B}=\emptyset$\;
 \For{$t=1$ to $K$}
 {
 compute the actual set of actions: $\mathcal{A}=\mathcal{C}\setminus\mathcal{B}$\;
 simulate one step with the $\epsilon$-greedy policy $\pi_{\epsilon}$:\\
 $a=\pi_{\epsilon}(\bm{s})$; $\bm{s}^{\prime},r=\text{STEP}(\bm{s},a)$\;
 $\mathcal{B}\leftarrow\mathcal{B}\cup\{a\}$\;
 add the experience $(\bm{s},a,r,\bm{s}^{\prime})$ into $\mathcal{M}$\;
 $\bm{s}\leftarrow\bm{s}^{\prime}$\;
 }
 randomly sample a mini-batch $\mathcal{B}$ from $\mathcal{M}$\;
 \For{all $(\bm{s},a,r,\bm{s}^{\prime})\in\mathcal{B}$}
 {
 calculate the learning target according to Eq.~(\ref{eq:target}):\\
 $y=r+\gamma\max_{a^{\prime}}Q(\bm{s}^{\prime},a^{\prime};\bm{w})$\;
 }
 carry out a gradient descent step on $\mathcal{L}$ w.r.t. $\bm{w}$ according to Eq.~(\ref{eq:sgd}):\\
 $\nabla_{\bm{w}}\mathcal{L}=\mathbb{E}_{(\bm{s},a,r,\bm{s}^{\prime})}[(y-Q(\bm{s},a;\bm{w}))]\nabla_{\bm{w}}Q(\bm{s},a;\bm{w})$\;
 update Q-network weights.
 }
 \caption{Training}
\end{algorithm}

\textbf{Q-network learning.} The Q-network is learned by minimizing the following mean squared Bellman error:
\begin{equation}
\label{eq:loss}
\mathcal{L}=\mathbb{E}_{(\bm{s},a,r,\bm{s}^{\prime})}[(y-\underbracket[0.5pt]{Q(\bm{s},a;\bm{w})}_{\text{Prediction}})^2] \,,
\end{equation}
where $\bm{w}$ represents network weights, and $y$ is the one-step ahead learning target:
\begin{equation}
\label{eq:target}
y=r+\gamma\max_{a^{\prime}}Q(\bm{s}^{\prime},a^{\prime};\bm{w}) \,.
\end{equation}
From Eq.~(\ref{eq:target}), it can be seen that the target is composed of the immediate reward $r$ and a discounted future reward. Ideally, the prediction of the current action-selection policy is supposed to be very close to the target, i.e., we want the error decrease. Hence we carry out a gradient descent step on $\mathcal{L}$ w.r.t. $\bm{w}$ according to:
\begin{equation}
\label{eq:sgd}
\nabla_{\bm{w}}\mathcal{L}=\mathbb{E}_{(\bm{s},a,r,\bm{s}^{\prime})}[(y-Q(\bm{s},a;\bm{w}))]\nabla_{\bm{w}}Q(\bm{s},a;\bm{w}) \,.
\end{equation}
\par

Actually, the learning of the Q-network for estimating the action-value function tends to be unstable. Therefore, in deep Q-learning, several techniques are used to address this problem, and they are detailed below.
\par
\textbf{Experience replay.} Here an experience refers to a 4-tuple $(\bm{s},a,r,\bm{s}^{\prime})$. Consecutively generated experiences in our model are highly correlated with each other, and this could result in unstable and inefficient learning that is also a  notorious problem in Q-learning. One solution to make the learning converge is to collect and store experiences in a replay memory, and during the training phase of the Q-network, mini-batches are randomly taken out from this replay memory and utilized for the Q-network training. This method has the following advantages:
\begin{itemize}
  \item One experience can be potentially used for many gradient descent steps, which improves data efficiency.
  \item Randomizing experiences breaks correlations among consecutive samples and therefore reduces the variance of gradient descent steps and stabilizes the learning of the network.
\end{itemize}
\par
\textbf{Exploration-exploitation.} To train the Q-network, we use an $\epsilon$-greedy policy, which means the agent either chooses actions at will with a probability $\epsilon$ or takes the best actions relying on the already learned band-selection policy with a probability $1-\epsilon$. The learning of the Q-network starts with a relatively large $\epsilon$ and then gradually decays it. The main idea behind this policy is that the agent is encouraged to try as many actions (i.e., various band combinations) as possible to begin with before it starts to see patterns. When it does not select actions at random, given a state, the agent is able to estimate the reward for each action. Thus the best action leading to the highest reward can be picked. Moreover, note that the $\epsilon$-greedy policy of our model is carried out on the actual action set $\mathcal{A}$, instead of the complete action set $\mathcal{C}$.

\begin{algorithm}[t]
\label{algo2}
\SetAlgoLined
 \textbf{function} $\bm{s}^{\prime},r=\text{STEP}(\bm{s},a)$ \textbf{:}\\
 get $\bm{s}^{\prime}$ based on $\bm{s}$ and $a$\;
 \eIf{$\bm{s}$ is $\vec{0}$}
 {
 $r=-\sum_{n=1}^NP(x_a^n)\log_2P(x_a^n)$\;
 }{
 calculate $r$ according to Eq.~(\ref{eq:2}):\\
 $r=\text{MIE}(\bm{s}^{\prime})-\text{MIE}(\bm{s})$\;
 }
 \caption{Environment simulation (based on information entropy)}
\end{algorithm}

\begin{table}[t!]
\caption{Illustration of the Q-network we use. Taking the Indian Pines dataset as an example.}
\label{tab:qnet_archi}
\centering
\linespread{1.}\selectfont
\begin{tabular}{llllll}
\Xhline{2\arrayrulewidth}
\textbf{Layer} & \textbf{Input} & \textbf{Output} & \textbf{\#Units} & \textbf{Connected to} & \textbf{Activation} \\
\hline
fc1 & (200,) & (400,) & 400 & input & ReLU \\
fc2 & (400,) & (400,) & 400 & fc1 & ReLU \\
fc3 & (400,) & (200,) & 200 & fc2 & linear \\
\Xhline{2\arrayrulewidth}
\end{tabular}
\end{table}

\subsection{Implementation Details}
\label{subsec:imp}
In this work, we set the maximum size of the replay memory as 50000 and make use of a batch size of 100. The $\epsilon$-greedy policy starts with $\epsilon=1$ and decreases until $\epsilon=0.01$ in steps of $0.95$. The weights of the Q-network are initialized randomly from a uniform distribution~\cite{GlorotB10}, and we note that outcomes are not sensitive to this initialization. For training the Q-network, Nesterov Adam~\cite{nadam2} is chosen as the optimizer, and its parameters are set as recommended, i.e., $\beta_1=0.9$ and $\beta_2=0.999$. In addition, a learning rate of $1\times10^{-4}$ is used. The training procedure and environment simulation of our model are shown in Algorithm~\ref{algo1} and~\ref{algo2}. Once the agent is trained with Algorithm~\ref{algo1}, it learns a band-selection policy that guides the agent to choose the band with the maximum estimated Q-value at each step. It should be noted that no matter in the training or the test phase, the agent is supposed to select spectral bands without duplicates in an episode.

\begin{figure}[!t]
\centering
\includegraphics[width=\columnwidth]{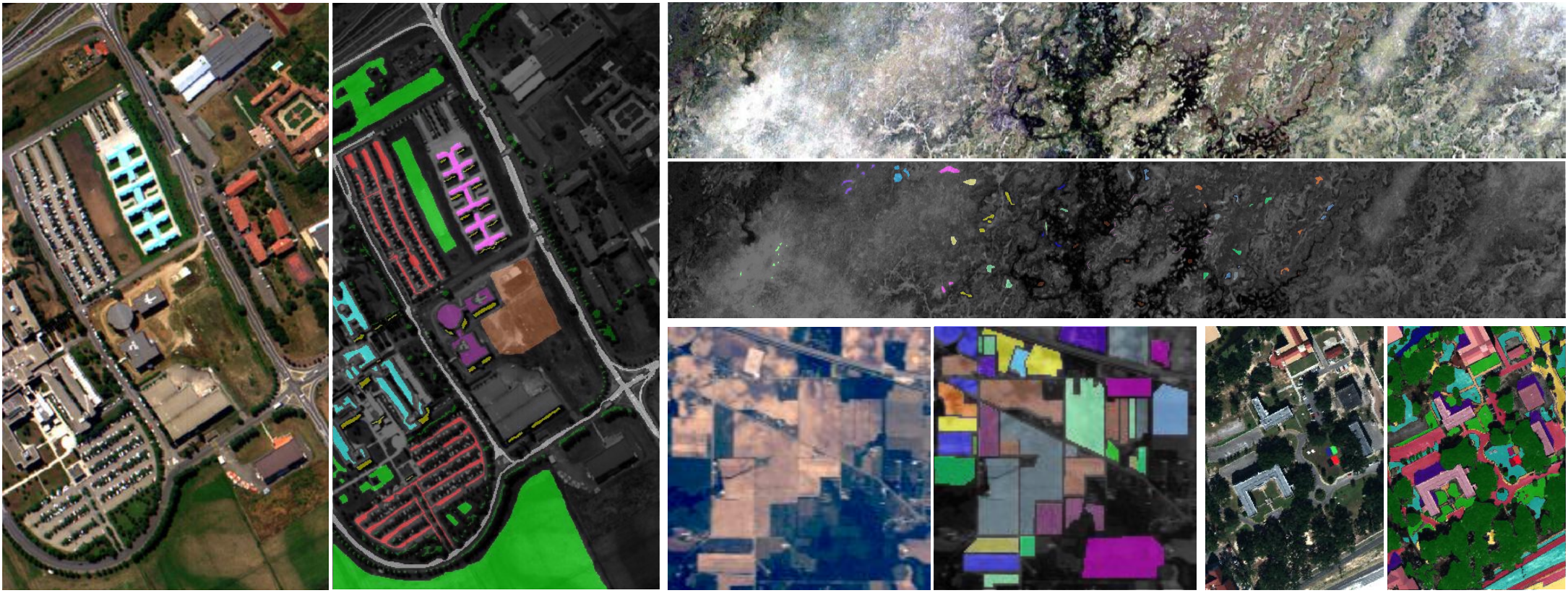}
\renewcommand{\figurename}{Fig.}
\caption{\label{fig:datasets} From left to right and top to bottom: True-color composite images and ground truth data of the Pavia University, Botswana, Indian Pines, and MUUFL Gulfport Data Sets.}
\end{figure}

\section{Experiments and Analysis}
\label{sec:exp}
\subsection{Hyperspectral Data Set Description}
\subsubsection{Indian Pines}
This scene was collected in June 1992 via NASA/JPL's Airborne Visible/Infrared Imaging Spectrometer (AVIRIS) sensor. It covers a geographical area in Northwestern Indiana, United States. This data set includes 145$\times$145 pixels, and its spatial resolution is 20 m/pixel. There are totally 220 spectral bands, and their wavelength values range between 400 nm and 2500 nm. The ground truth provided by the data set involves 16 classes of interest, of which the majority of these classes are related to crops at variant growth stages (cf.~Fig.~\ref{fig:datasets}). Before performing band selection algorithms, we remove 20 bands, i.e., 104-108, 150-163, and 220, as they are both water absorption ones, and as a result, 200 spectral bands are eventually used in total.

\subsubsection{Pavia University}
The second scene was captured through Reflective Optics Spectrographic Imaging System (ROSIS) on an aircraft operated by the German Aerospace Center (DLR) in 2002. It covers an area of the University of Pavia and is composed of 103 spectral bands in the wavelength range of 430-860 nm after discarding 12 noisy bands and 610$\times$340 pixels. The spatial resolution of this scene is 1.3 m/pixel. Except for unknown pixels, 9 land cover categories are labeled manually in the ground truth. Fig.~\ref{fig:datasets} exhibits the true-color composite image and reference data of the Pavia University data set.

\subsubsection{Botswana}
The Botswana scene was collected by a hyperspectral sensor, Hyperion, on NASA EO-1 satellite in May 2001. It covers a 7.7 km strip in the Okavango Delta, Botswana and includes 1476$\times$256 pixels. Its spatial resolution is 30 m/pixel. 242 spectral bands whose wavelength varies between 400 nm and 2500 nm are originally captured in this data set, but only 145 bands are used in our study after we remove noisy and uncalibrated bands. There are 14 classes representing different land covers included in the ground truth provided by the data set (see Fig.~\ref{fig:datasets}).

\begin{figure}[t]
\centering
\includegraphics[width=0.6\columnwidth]{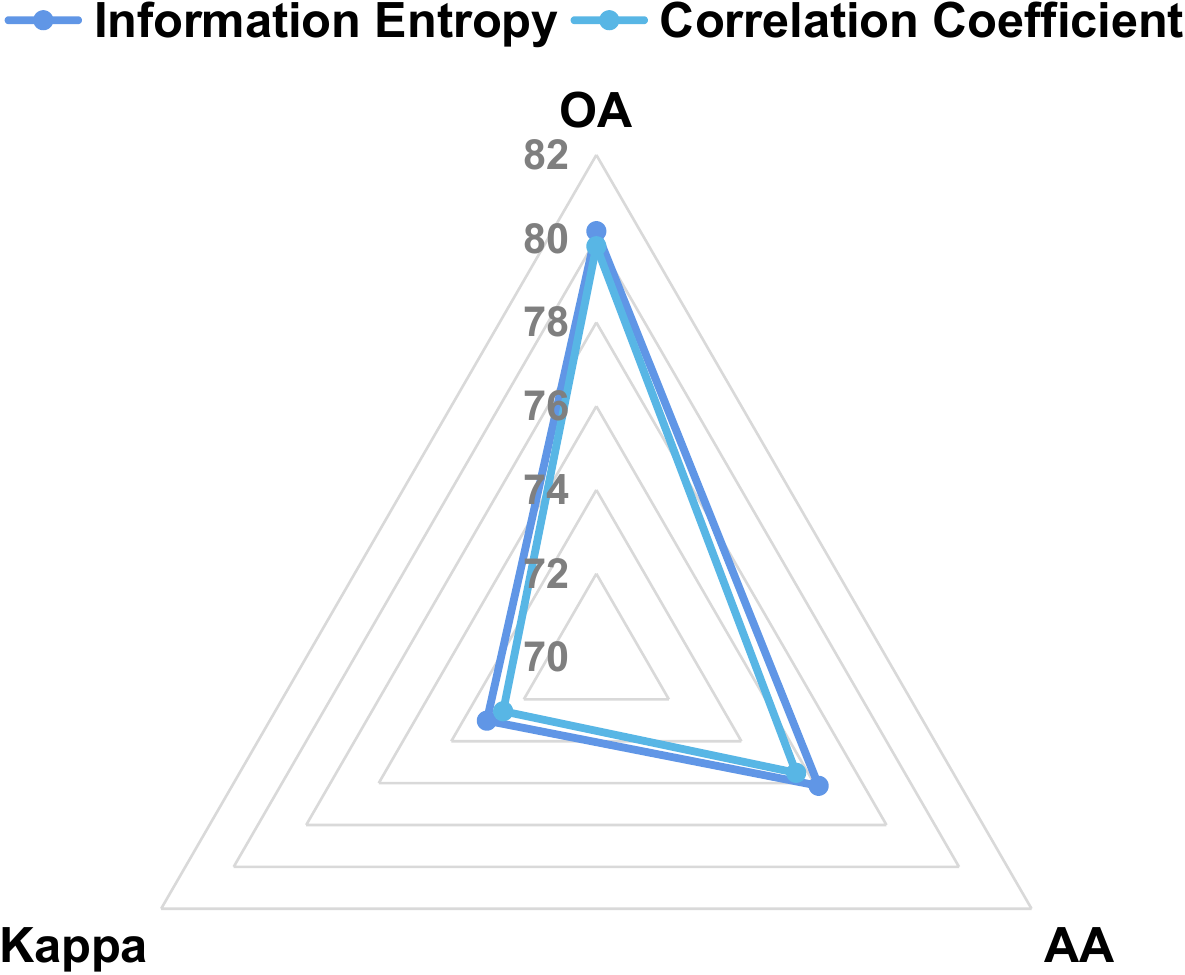}
\renewcommand{\figurename}{Fig.}
\caption{\label{fig:IEvsCorr} Comparison of two reward schemes, namely information entropy and correlation coefficient, on the Pavia University data set.}
\end{figure}

\subsubsection{MUUFL Gulfport}
The MUUFL Gulfport data set~\cite{gulfport1,gulfport2} was acquired over the campus of the University of Southern Mississippi Gulf Park, Long Beach, Mississippi in 2010. It includes co-registered hyperspectral and LiDAR data, but in this work, we only use the hyperspectral data that originally contains 72 bands. However, due to noise, the first four and last four bands are omitted, bringing about an image with 64 bands. There is a total of 325$\times$337 pixels, and the provided ground truth map includes 11 classes. This data set can be used to evaluate the performance of different band selection methods under the circumstance where a hyperspectral data set has a limited number of bands.
\par
\RR{}{Table~\ref{tab:datasets_statistic} outlines the number of labeled samples and classes of each data set.}

\begin{table*}[t!]
\caption{Number of Labeled Samples in the Indian Pines, Pavia University, Botswana, and MUUFL Gulfport Data Sets.}
\label{tab:datasets_statistic}
\centering
\linespread{1.}\selectfont
\begin{tabular}{r|cc|cc|cc|cc}
\Xhline{2\arrayrulewidth}
\multirow{2}{0.3cm}{\textbf{No.}} & \multicolumn{2}{c|}{\textbf{Indian Pines}} & \multicolumn{2}{c|}{\textbf{Pavia University}} & \multicolumn{2}{c}{\textbf{Botswana}} & \multicolumn{2}{c}{\textbf{MUUFL Gulfport}} \\
\cline{2-9}
 & \textbf{Class name} & \textbf{\# Samples} & \textbf{Class name} & \textbf{\# Samples} & \textbf{Class name} & \textbf{\# Samples} & \textbf{Class name} & \textbf{\# Samples} \\
\hline
\hline
1 & Corn-notill & 1434 & Asphalt & 6631 &  Water & 270 &  Trees & 23246 \\
2 & Corn-min & 834 & Meadows & 18649 &  Hippo grass & 101 &   Mostly grass & 4270 \\
3 & Corn & 234 & Gravel & 2099 & Floodplain grasses 1 & 251 & Mixed ground surface & 6882 \\
4 & Grass-pasture & 497 & Trees & 3064 & Floodplain grasses 2 & 215 & Dirt/Sand & 1826 \\
5 & Grass-trees & 747 & Metal sheets & 1345 & Reeds 1 & 269 & Road & 6687 \\
6 & Hay-windrowed & 489 & Bare Soil & 5029 & Riparian & 269 & Water & 466 \\
7 & Soybean-notill & 968 & Bitumen & 1330 & Firescar 2 & 259 & Building shadow & 2233 \\
8 & Soybean-mintill & 2468 & Bricks & 3682 & Island interior & 203 & Buildings & 6240 \\
9 & Soybean-clean & 614 & Shadows & 947 & Acacia woodlands & 314 & Sidewalk & 1385 \\
10 & Wheat & 212 & - & - & Acacia shrublands & 248 & Yellow curb & 183 \\
11 & Woods & 1294 & - & - & Acacia grasslands & 305 & Cloth panels & 269 \\
12 & Buildings-grass-trees & 380 & - & - & Short mopane & 181 & - & - \\
13 & Stone-steel-towers & 95 & - & - & Mixed mopane & 268 & - & - \\
14 & Alfalfa & 54 & - & - & Exposed soils & 95 & - & - \\
15 & Grass-pasture-mowed & 26 & - & - & - & - & - & - \\
16 & Oats & 20 & - & - & - & - & - & - \\
\Xhline{2\arrayrulewidth}
\end{tabular}
\end{table*}

\subsection{Experiment Setting}
\label{sunsec:exp-b}
\subsubsection{Evaluation}
We use classification tasks to validate the effectiveness of selected bands. As to evaluation measurements, we make use of the following ones:
\begin{itemize}
  \item Overall accuracy (OA): This metric is calculated by summing the amount of correctly identified data and dividing by the total amount of data.
  \item Average accuracy (AA): This measurement is computed by averaging all per-class accuracies.
  \item Kappa coefficient: This coefficient evaluates the agreement between predictions and labels.
\end{itemize}

\subsubsection{Band Selection Methods in Comparison}
To evaluate the proposed approach, we compare it with several state-of-the-art band selection algorithms that are listed as follows:
\begin{itemize}
  \item MVPCA~\cite{ChangDSA99}: A ranking-based band selection method that uses an eigenanalysis-based criterion to prioritize spectral bands.
  \item ICA~\cite{DuQWRS03}: A band selection approach that compares mean absolute independent component analysis (ICA) coefficients of individual spectral bands and picks independent ones including the maximum information.
  \item IE~\cite{XieJARS17}: A ranking-based band selection algorithm in which band priority is calculated based on information entropy.
  \item MEV-SFS~\cite{ZhangLDZ18}: A searching-based band selection method that combines maximum ellipsoid volume (MEV)~\cite{Sheffield85} method with sequential forward search (SFS). The MEV model deems an optimal band subset as a band combination with the maximum volume.
  \item OPBS~\cite{ZhangLDZ18}: An accelerated version of MEV-SFS that takes advantage of a relationship between orthogonal projections (OPs) and the ellipsoid volume of bands to find out an optimal band combination.
  \item WaLuDi~\cite{UsoPSG07}: A hierarchical clustering-based band selection method that uses Kullback-Leibler divergence as the criterion of the clustering algorithm.
  \item E-FDPC~\cite{JiaTZL16}: A clustering-based band selection approach that makes use of an enhanced version of fast density peak-based clustering (FDPC)~\cite{FDPC} algorithm by introducing an exponential learning rule and a parameter to control the weight between local density and intra-cluster similarity.
  \item OCF~\cite{WangZL18}: A band selection method using an optimal clustering algorithm that is capable of achieving optimal clustering results for an objective function with a carefully designed constraint.
  \item ASPS~\cite{WangLL19}: A clustering-based band selection method that exploits an adaptive subspace partition strategy.
  \item DARecNet~\cite{Roy2020}: An unsupervised convolutional neural network (CNN) for band selection tasks. It employs a dual-attention mechanism, i.e., spatial position attention and channel attention, to learn to reconstruct hyperspectral images. Once the network is trained, bands are selected according to entropies of the reconstructed bands.
  \item DRL: Our proposed deep reinforcement learning model for unsupervised hyperspectral band selection.
\end{itemize}

\subsubsection{Classification Setting}
We consider four commonly used classifiers in the remote sensing community to implement hyperspectral image classification. They are as follows:
\begin{itemize}
  \item k-NN: A k-nearest neighbors algorithm (the number of neighbors is set to 3).
  \item RF: A random forest being made up of 200 decision trees.
  \item MLP: A multilayer perceptron that consists of three fully connected layers. The first two layers contain 256 units, and their outputs are activated by Leaky RuLU. For the last layer, the number of units equals the number of classes, and the used activation function is softmax. In the learning phase, we select Adam as the optimizer and define the loss function as categorical cross-entropy. The learning rate is set to 0.0005, and the training epochs is 2000 for the purpose of sufficient learning.
  \item SVM-RBF\footnote{\url{https://www.csie.ntu.edu.tw/~cjlin/libsvmtools/}}: A support vector machine (SVM) equipped with radial basis function (RBF) kernel. A five-fold cross-validation method is utilized to determine optimal hyper-parameters, i.e., $\gamma$ and $C$.
\end{itemize}

For both the Indian Pines and the Botswana data sets, we randomly select 10\% samples from each class as training instances, while the remaining are exploited to test models. Regarding the Pavia University and MUUFL Gulfport data sets, 1\% samples per class are chosen randomly to build the training set, and all the other samples are utilized for the purpose of testing. In order to know the stability of various band selection models, final results are achieved by averaging 10 individual runs, and we report the mean and standard deviation of performance metrics of different approaches over the 10 runs.

\begin{table*}[t!]
\scriptsize\addtolength{\tabcolsep}{-3.9pt}
\caption{Comparisons in quantitative metrics among different band selection methods on the \underline{Indian Pines} data set with 30 selected bands. We report the mean and standard deviation of performance metrics of different approaches over 10 individual runs. The best classification performance is highlighted in \textbf{bold}.}
\label{tab:indian_pines}
\centering
\linespread{1.}\selectfont
\begin{tabular}{r|ccc|ccc|ccc|ccc}
\Xhline{2\arrayrulewidth}
\multirow{2}{1cm}{\textbf{Models}} & \multicolumn{3}{c|}{\textbf{k-NN}} & \multicolumn{3}{c|}{\textbf{RF}} & \multicolumn{3}{c|}{\textbf{MLP}} & \multicolumn{3}{c}{\textbf{SVM-RBF}} \\
\cline{2-13}
 & \textbf{OA} & \textbf{AA} & \textbf{Kappa} & \textbf{OA} & \textbf{AA} & \textbf{Kappa} & \textbf{OA} & \textbf{AA} & \textbf{Kappa} & \textbf{OA} & \textbf{AA} & \textbf{Kappa} \\
\hline
\hline
MVPCA~\cite{ChangDSA99} & 59.45$\pm$0.95 & 48.86$\pm$2.22 & 53.58$\pm$1.11 & 64.50$\pm$0.52 & 55.15$\pm$1.43 & 59.04$\pm$0.58 & 61.85$\pm$2.57 & 56.38$\pm$2.18 & 56.15$\pm$2.67 & 70.10$\pm$1.14 & 62.33$\pm$2.83 & 65.65$\pm$1.32 \\
ICA~\cite{DuQWRS03} & 65.80$\pm$0.60 & 57.72$\pm$1.34 & 60.91$\pm$0.70 & 71.30$\pm$0.75 & 60.90$\pm$1.26 & 66.90$\pm$0.90 & 70.92$\pm$1.58 & 64.51$\pm$2.51 & 66.57$\pm$2.06 & 73.85$\pm$1.18 & 67.75$\pm$1.94 & 69.97$\pm$1.38 \\
IE~\cite{XieJARS17} & 59.69$\pm$0.71 & 51.31$\pm$1.36 & 53.92$\pm$0.82 & 63.82$\pm$0.42 & 54.84$\pm$1.51 & 58.25$\pm$0.44 & 61.83$\pm$1.18 & 56.64$\pm$1.54 & 55.99$\pm$1.45 & 70.42$\pm$0.63 & 62.46$\pm$1.86 & 66.08$\pm$0.69 \\
MEV-SFS~\cite{ZhangLDZ18} & 60.71$\pm$0.78 & 52.91$\pm$1.99 & 55.15$\pm$0.86 & 69.21$\pm$0.76 & 57.96$\pm$1.18 & 64.35$\pm$0.88 & 67.11$\pm$1.55 & 60.65$\pm$2.32 & 62.18$\pm$1.93 & 69.72$\pm$0.94 & 63.40$\pm$2.08 & 65.23$\pm$1.14 \\
OPBS~\cite{ZhangLDZ18} & 63.17$\pm$0.58 & 54.86$\pm$1.17 & 57.90$\pm$0.67 & 71.08$\pm$0.70 & 59.93$\pm$1.21 & 66.61$\pm$0.81 & 69.30$\pm$0.89 & 62.58$\pm$1.89 & 64.75$\pm$1.07 & 71.08$\pm$0.80 & 63.88$\pm$1.78 & 66.86$\pm$0.94 \\
WaLuDi~\cite{UsoPSG07} & 63.27$\pm$0.68 & 52.77$\pm$1.98 & 57.97$\pm$0.77 & 73.18$\pm$0.76 & 58.97$\pm$1.70 & 69.01$\pm$0.92 & 73.36$\pm$1.21 & 65.45$\pm$2.19 & 69.54$\pm$1.42 & 77.52$\pm$0.61 & 69.91$\pm$1.73 & 74.25$\pm$0.68 \\
E-FDPC~\cite{JiaTZL16} & 61.83$\pm$0.86 & 48.52$\pm$1.18 & 56.21$\pm$0.96 & 64.09$\pm$0.50 & 47.72$\pm$1.17 & 58.22$\pm$0.60 & 62.67$\pm$1.21 & 46.75$\pm$1.54 & 56.44$\pm$1.34 & 67.24$\pm$0.69 & 56.86$\pm$3.22 & 62.06$\pm$0.74 \\
OCF~\cite{WangZL18} & 63.70$\pm$0.92 & 53.63$\pm$1.88 & 58.45$\pm$1.07 & 73.41$\pm$1.01 & 59.44$\pm$1.63 & 69.31$\pm$1.21 & 74.06$\pm$1.02 & 67.35$\pm$1.98 & 70.35$\pm$1.19 & 78.13$\pm$0.64 & 71.45$\pm$1.66 & 74.98$\pm$0.70 \\
ASPS~\cite{WangLL19} & 62.92$\pm$0.49 & 52.20$\pm$1.24 & 57.51$\pm$0.58 & 73.13$\pm$0.87 & 59.11$\pm$1.60 & 68.96$\pm$1.04 & 73.49$\pm$0.50 & 67.19$\pm$2.71 & 69.62$\pm$0.61 & 77.13$\pm$0.95 & 71.37$\pm$2.83 & 73.78$\pm$1.12 \\
DARecNet~\cite{Roy2020} & 64.21$\pm$0.82 & 57.72$\pm$1.71 & 59.10$\pm$0.93 & 71.18$\pm$0.60 & 60.65$\pm$1.52 & 66.85$\pm$0.70 & 70.24$\pm$0.70 & 65.08$\pm$2.14 & 65.96$\pm$0.78 & 74.90$\pm$0.74 & 69.46$\pm$2.56 & 71.25$\pm$0.85 \\
\hline
DRL (proposed) & \textbf{67.85$\pm$0.63} & \textbf{59.85$\pm$1.17} & \textbf{63.22$\pm$0.73} & \textbf{74.16$\pm$0.79} & \textbf{61.47$\pm$1.42} & \textbf{70.18$\pm$0.94} & \textbf{75.80$\pm$0.76} & \textbf{71.32$\pm$1.65} & \textbf{72.34$\pm$0.90} & \textbf{78.70$\pm$0.86} & \textbf{74.39$\pm$2.14} & \textbf{75.62$\pm$1.00} \\
\Xhline{2\arrayrulewidth}
\end{tabular}
\end{table*}

\begin{table*}[t!]
\scriptsize\addtolength{\tabcolsep}{-3.9pt}
\caption{Comparisons in quantitative metrics among different band selection methods on the \underline{Pavia University} data set with 30 selected bands. We report the mean and standard deviation of performance metrics of different approaches over 10 individual runs. The best classification performance is highlighted in \textbf{bold}.}
\label{tab:pavia_uni}
\centering
\linespread{1.}\selectfont
\begin{tabular}{r|ccc|ccc|ccc|ccc}
\Xhline{2\arrayrulewidth}
\multirow{2}{1cm}{\textbf{Models}} & \multicolumn{3}{c|}{\textbf{k-NN}} & \multicolumn{3}{c|}{\textbf{RF}} & \multicolumn{3}{c|}{\textbf{MLP}} & \multicolumn{3}{c}{\textbf{SVM-RBF}} \\
\cline{2-13}
 & \textbf{OA} & \textbf{AA} & \textbf{Kappa} & \textbf{OA} & \textbf{AA} & \textbf{Kappa} & \textbf{OA} & \textbf{AA} & \textbf{Kappa} & \textbf{OA} & \textbf{AA} & \textbf{Kappa} \\
\hline
\hline
MVPCA~\cite{ChangDSA99} & 75.76$\pm$1.07 & 68.52$\pm$1.42 & 67.05$\pm$1.42 & 78.83$\pm$0.80 & 70.23$\pm$1.96 & 71.26$\pm$1.13 & 78.00$\pm$0.56 & 69.44$\pm$2.14 & 70.19$\pm$0.86 & 85.43$\pm$0.99 & 77.52$\pm$2.85 & 80.44$\pm$1.35 \\
ICA~\cite{DuQWRS03} & 72.41$\pm$0.88 & 66.39$\pm$1.29 & 62.53$\pm$1.16 & 75.19$\pm$0.95 & 67.83$\pm$1.11 & 65.85$\pm$1.05 & 75.99$\pm$1.02 & 69.93$\pm$1.28 & 67.68$\pm$1.37 & 80.25$\pm$1.19 & 74.19$\pm$2.53 & 73.23$\pm$1.87 \\
IE~\cite{XieJARS17} & 76.29$\pm$1.08 & 68.95$\pm$1.61 & 67.75$\pm$1.44 & 79.08$\pm$0.72 & 69.98$\pm$1.99 & 71.49$\pm$1.06 & 78.47$\pm$0.94 & 70.75$\pm$2.33 & 70.93$\pm$1.20 & 85.74$\pm$1.10 & 77.75$\pm$3.51 & 80.90$\pm$1.54 \\
MEV-SFS~\cite{ZhangLDZ18} & 78.25$\pm$0.76 & 73.63$\pm$1.30 & 70.44$\pm$0.98 & 81.40$\pm$0.77 & 75.38$\pm$0.90 & 74.56$\pm$0.93 & 83.01$\pm$0.61 & 79.42$\pm$1.21 & 77.19$\pm$0.85 & 87.51$\pm$1.01 & 83.67$\pm$1.74 & 83.28$\pm$1.41 \\
OPBS~\cite{ZhangLDZ18} & 78.25$\pm$0.76 & 73.63$\pm$1.30 & 70.44$\pm$0.98 & 81.42$\pm$0.86 & 75.39$\pm$0.89 & 74.58$\pm$1.07 & 82.76$\pm$0.92 & 78.55$\pm$1.08 & 76.84$\pm$1.13 & 87.51$\pm$1.01 & 83.67$\pm$1.74 & 83.28$\pm$1.41 \\
WaLuDi~\cite{UsoPSG07} & 78.30$\pm$0.63 & 73.69$\pm$1.05 & 70.53$\pm$0.76 & 80.45$\pm$0.63 & 74.15$\pm$1.61 & 73.27$\pm$0.91 & 82.93$\pm$0.82 & 78.99$\pm$1.49 & 77.12$\pm$1.03 & 87.19$\pm$1.01 & 82.99$\pm$1.96 & 82.86$\pm$1.38 \\
E-FDPC~\cite{JiaTZL16} & 78.30$\pm$0.79 & 74.55$\pm$1.03 & 70.56$\pm$1.05 & 79.70$\pm$0.82 & 74.03$\pm$1.83 & 72.33$\pm$1.07 & 80.56$\pm$1.15 & 74.92$\pm$2.08 & 73.70$\pm$1.38 & 84.22$\pm$0.96 & 80.72$\pm$1.38 & 78.64$\pm$1.31 \\
OCF~\cite{WangZL18} & 78.07$\pm$0.82 & 73.34$\pm$1.44 & 70.21$\pm$1.12 & 80.58$\pm$0.77 & 74.25$\pm$1.38 & 73.47$\pm$1.02 & 82.00$\pm$0.86 & 77.77$\pm$1.32 & 75.77$\pm$1.03 & 86.61$\pm$0.82 & 83.58$\pm$0.95 & 82.09$\pm$1.06 \\
ASPS~\cite{WangLL19} & 78.23$\pm$0.59 & 73.80$\pm$1.04 & 70.42$\pm$0.79 & 80.93$\pm$0.54 & 74.87$\pm$1.33 & 73.96$\pm$0.72 & 82.70$\pm$1.17 & 79.19$\pm$1.19 & 76.79$\pm$1.50 & 87.07$\pm$0.92 & 83.53$\pm$1.83 & 82.67$\pm$1.27 \\
DARecNet~\cite{Roy2020} & 74.11$\pm$0.82 & 68.21$\pm$1.58 & 64.78$\pm$0.96 & 74.58$\pm$0.98 & 66.71$\pm$1.85 & 65.11$\pm$1.25 & 74.05$\pm$1.28 & 65.38$\pm$2.24 & 64.63$\pm$1.42 & 79.46$\pm$0.68 & 74.30$\pm$1.44 & 71.80$\pm$0.90 \\
\hline
DRL (proposed) & \textbf{80.18$\pm$0.74} & \textbf{76.13$\pm$1.30} & \textbf{73.02$\pm$0.91} & \textbf{81.78$\pm$0.64} & \textbf{75.41$\pm$1.61} & \textbf{75.12$\pm$0.94} & \textbf{84.73$\pm$0.91} & \textbf{80.74$\pm$2.01} & \textbf{79.45$\pm$1.36} & \textbf{89.05$\pm$0.61} & \textbf{85.85$\pm$1.17} & \textbf{85.35$\pm$0.83} \\
\Xhline{2\arrayrulewidth}
\end{tabular}
\end{table*}

\begin{table*}[t!]
\scriptsize\addtolength{\tabcolsep}{-3.9pt}
\caption{Comparisons in quantitative metrics among different band selection methods on the \underline{Botswana} data set with 30 selected bands. We report the mean and standard deviation of performance metrics of different approaches over 10 individual runs. The best classification performance is highlighted in \textbf{bold}.}
\label{tab:botswana}
\centering
\linespread{1.}\selectfont
\begin{tabular}{r|ccc|ccc|ccc|ccc}
\Xhline{2\arrayrulewidth}
\multirow{2}{1cm}{\textbf{Models}} & \multicolumn{3}{c|}{\textbf{k-NN}} & \multicolumn{3}{c|}{\textbf{RF}} & \multicolumn{3}{c|}{\textbf{MLP}} & \multicolumn{3}{c}{\textbf{SVM-RBF}} \\
\cline{2-13}
 & \textbf{OA} & \textbf{AA} & \textbf{Kappa} & \textbf{OA} & \textbf{AA} & \textbf{Kappa} & \textbf{OA} & \textbf{AA} & \textbf{Kappa} & \textbf{OA} & \textbf{AA} & \textbf{Kappa} \\
\hline
\hline
MVPCA~\cite{ChangDSA99} & 81.43$\pm$1.15 & 82.80$\pm$1.01 & 79.88$\pm$1.24 & 80.24$\pm$1.64 & 81.04$\pm$1.70 & 78.58$\pm$1.77 & 82.98$\pm$0.99 & 83.04$\pm$0.90 & 81.55$\pm$1.07 & 87.74$\pm$1.02 & 88.95$\pm$0.91 & 86.72$\pm$1.11 \\
ICA~\cite{DuQWRS03} & 82.89$\pm$0.85 & 83.45$\pm$1.04 & 81.46$\pm$0.93 & 81.96$\pm$1.10 & 82.77$\pm$1.09 & 80.45$\pm$1.19 & 87.21$\pm$0.77 & 87.51$\pm$0.79 & 86.14$\pm$0.83 & 88.58$\pm$1.05 & 89.36$\pm$1.21 & 87.62$\pm$1.14 \\
IE~\cite{XieJARS17} & 79.44$\pm$0.86 & 80.01$\pm$0.92 & 77.72$\pm$0.93 & 78.55$\pm$1.29 & 78.48$\pm$1.31 & 76.74$\pm$1.39 & 85.59$\pm$1.05 & 86.15$\pm$1.03 & 84.39$\pm$1.13 & 88.51$\pm$1.29 & 89.50$\pm$1.25 & 87.55$\pm$1.40 \\
MEV-SFS~\cite{ZhangLDZ18} & 84.33$\pm$0.92 & 85.53$\pm$0.85 & 83.03$\pm$0.99 & 84.01$\pm$0.90 & 84.93$\pm$0.88 & 82.68$\pm$0.98 & 87.72$\pm$0.40 & 87.80$\pm$0.57 & 86.69$\pm$0.43 & 90.29$\pm$1.03 & 91.26$\pm$1.07 & 89.49$\pm$1.12 \\
OPBS~\cite{ZhangLDZ18} & 84.33$\pm$0.92 & 85.53$\pm$0.85 & 83.03$\pm$0.99 & 84.20$\pm$0.83 & 85.38$\pm$0.98 & 82.88$\pm$0.91 & 87.70$\pm$0.63 & 87.89$\pm$0.84 & 86.67$\pm$0.68 & 90.29$\pm$1.03 & 91.26$\pm$1.07 & 89.49$\pm$1.12 \\
WaLuDi~\cite{UsoPSG07} & 86.49$\pm$0.84 & 87.82$\pm$0.80 & 85.36$\pm$0.91 & 85.16$\pm$0.85 & 85.90$\pm$0.90 & 83.92$\pm$0.92 & 88.88$\pm$0.81 & \textbf{89.20$\pm$1.03} & 87.95$\pm$0.88 & 90.69$\pm$0.98 & 91.66$\pm$1.01 & 89.92$\pm$1.06 \\
E-FDPC~\cite{JiaTZL16} & 79.73$\pm$1.43 & 80.62$\pm$1.16 & 78.05$\pm$1.54 & 79.05$\pm$1.07 & 79.68$\pm$1.29 & 77.30$\pm$1.16 & 74.45$\pm$1.01 & 73.60$\pm$1.46 & 72.32$\pm$1.10 & 83.90$\pm$1.21 & 84.54$\pm$1.37 & 82.55$\pm$1.32 \\
OCF~\cite{WangZL18} & 84.76$\pm$1.12 & 86.02$\pm$0.93 & 83.50$\pm$1.21 & 83.66$\pm$0.73 & 84.72$\pm$0.89 & 82.31$\pm$0.79 & 86.64$\pm$1.02 & 86.83$\pm$1.23 & 85.53$\pm$1.10 & 89.64$\pm$1.25 & 90.66$\pm$1.20 & 88.77$\pm$1.35 \\
ASPS~\cite{WangLL19} & 85.60$\pm$0.75 & 86.96$\pm$0.63 & 84.41$\pm$0.81 & 84.41$\pm$0.95 & 85.20$\pm$0.88 & 83.11$\pm$1.02 & 87.48$\pm$0.84 & 87.17$\pm$1.34 & 86.43$\pm$0.92 & 90.45$\pm$0.90 & 91.42$\pm$0.74 & 89.65$\pm$0.98 \\
DARecNet~\cite{Roy2020} & 81.52$\pm$0.78 & 83.21$\pm$1.05 & 79.98$\pm$0.85 & 80.99$\pm$0.98 & 82.53$\pm$1.01 & 79.41$\pm$1.06 & 83.08$\pm$1.14 & 83.65$\pm$1.15 & 81.67$\pm$1.24 & 88.14$\pm$0.66 & 89.25$\pm$0.58 & 87.15$\pm$0.71 \\
\hline
DRL (proposed) & \textbf{86.83$\pm$0.70} & \textbf{88.28$\pm$0.70} & \textbf{85.74$\pm$0.76} & \textbf{86.34$\pm$0.84} & \textbf{87.36$\pm$0.93} & \textbf{85.20$\pm$0.91} & \textbf{88.98$\pm$0.68} & 88.87$\pm$0.86 & \textbf{88.06$\pm$0.74} & \textbf{92.14$\pm$0.77} & \textbf{92.81$\pm$0.90} & \textbf{91.48$\pm$0.83} \\
\Xhline{2\arrayrulewidth}
\end{tabular}
\end{table*}

\begin{table*}[t!]
\scriptsize\addtolength{\tabcolsep}{-3.9pt}
\caption{Comparisons in quantitative metrics among different band selection methods on the \underline{MUUFL Gulfport} data set with 30 selected bands. We report the mean and standard deviation of performance metrics of different approaches over 10 individual runs. The best classification performance is highlighted in \textbf{bold}.}
\label{tab:muufl_gulfport}
\centering
\linespread{1.}\selectfont
\begin{tabular}{r|ccc|ccc|ccc|ccc}
\Xhline{2\arrayrulewidth}
\multirow{2}{1cm}{\textbf{Models}} & \multicolumn{3}{c|}{\textbf{k-NN}} & \multicolumn{3}{c|}{\textbf{RF}} & \multicolumn{3}{c|}{\textbf{MLP}} & \multicolumn{3}{c}{\textbf{SVM-RBF}} \\
\cline{2-13}
 & \textbf{OA} & \textbf{AA} & \textbf{Kappa} & \textbf{OA} & \textbf{AA} & \textbf{Kappa} & \textbf{OA} & \textbf{AA} & \textbf{Kappa} & \textbf{OA} & \textbf{AA} & \textbf{Kappa} \\
\hline
\hline
MVPCA~\cite{ChangDSA99} & 69.66$\pm$0.79 & 46.47$\pm$3.65 & 59.14$\pm$1.07 & 73.28$\pm$1.01 & 50.62$\pm$2.68 & 63.95$\pm$1.49 & 76.66$\pm$0.87 & 52.43$\pm$2.19 & 68.96$\pm$1.15 & 77.32$\pm$1.24 & 54.43$\pm$3.62 & 69.83$\pm$1.65 \\
ICA~\cite{DuQWRS03} & 78.77$\pm$0.89 & 58.45$\pm$2.27 & 71.81$\pm$1.16 & 80.24$\pm$0.56 & 58.70$\pm$2.89 & 73.89$\pm$0.69 & 82.01$\pm$0.42 & 66.33$\pm$2.04 & 76.23$\pm$0.59 & 81.72$\pm$0.85 & 61.00$\pm$3.87 & 75.81$\pm$1.09 \\
IE~\cite{XieJARS17} & 69.66$\pm$0.79 & 46.47$\pm$3.65 & 59.14$\pm$1.07 & 73.54$\pm$0.80 & 49.89$\pm$2.51 & 64.31$\pm$1.25 & 77.02$\pm$0.60 & 52.83$\pm$1.22 & 69.49$\pm$0.75 & 77.28$\pm$1.20 & 55.35$\pm$3.30 & 69.74$\pm$1.67 \\
MEV-SFS~\cite{ZhangLDZ18} & 77.61$\pm$0.85 & 58.25$\pm$3.42 & 70.17$\pm$1.17 & 79.92$\pm$0.54 & 60.39$\pm$1.81 & 73.43$\pm$0.69 & 81.55$\pm$0.77 & 64.30$\pm$1.20 & 75.52$\pm$0.88 & 81.24$\pm$0.98 & 62.40$\pm$3.36 & 75.14$\pm$1.24 \\
OPBS~\cite{ZhangLDZ18} & 77.61$\pm$0.85 & 58.25$\pm$3.42 & 70.17$\pm$1.17 & 80.08$\pm$0.62 & 59.17$\pm$2.73 & 73.66$\pm$0.81 & 81.88$\pm$0.74 & 64.94$\pm$1.77 & 76.01$\pm$1.01 & 81.24$\pm$0.98 & 62.40$\pm$3.36 & 75.14$\pm$1.24 \\
WaLuDi~\cite{UsoPSG07} & 79.26$\pm$0.70 & 59.38$\pm$2.55 & 72.47$\pm$0.96 & 80.52$\pm$0.56 & 59.85$\pm$2.12 & 74.26$\pm$0.69 & 83.10$\pm$0.56 & 67.69$\pm$1.37 & 77.60$\pm$0.79 & 82.62$\pm$0.88 & 65.48$\pm$4.38 & 77.01$\pm$1.16 \\
E-FDPC~\cite{JiaTZL16} & 78.91$\pm$0.78 & 58.28$\pm$1.67 & 71.96$\pm$1.11 & 80.30$\pm$0.55 & 60.66$\pm$1.93 & 73.94$\pm$0.76 & 82.76$\pm$0.73 & \textbf{69.42$\pm$1.88} & 77.17$\pm$0.98 & 83.10$\pm$0.75 & 67.58$\pm$2.66 & 77.66$\pm$1.01 \\
OCF~\cite{WangZL18} & 79.39$\pm$0.67 & 59.82$\pm$1.21 & 72.65$\pm$0.96 & 80.54$\pm$0.66 & 59.79$\pm$2.36 & 74.28$\pm$0.81 & 83.11$\pm$0.78 & 67.43$\pm$1.83 & 77.58$\pm$1.05 & 82.75$\pm$1.02 & 66.43$\pm$4.91 & 77.21$\pm$1.35 \\
ASPS~\cite{WangLL19} & 78.86$\pm$0.74 & 60.10$\pm$2.31 & 71.90$\pm$1.08 & 80.34$\pm$0.60 & 60.38$\pm$2.01 & 74.04$\pm$0.72 & 83.03$\pm$0.77 & 67.28$\pm$1.92 & 77.53$\pm$1.04 & 82.73$\pm$0.84 & 67.32$\pm$2.50 & 77.15$\pm$1.09 \\
DARecNet~\cite{Roy2020} & 76.34$\pm$0.86 & 56.82$\pm$1.79 & 68.51$\pm$1.20 & 79.44$\pm$0.93 & 59.94$\pm$2.17 & 72.78$\pm$1.22 & 80.89$\pm$1.20 & 60.25$\pm$3.10 & 74.57$\pm$1.68 & 80.60$\pm$0.87 & 60.35$\pm$4.45 & 74.27$\pm$1.15 \\
\hline
DRL (proposed) & \textbf{79.68$\pm$0.63} & \textbf{60.85$\pm$2.03} & \textbf{73.08$\pm$0.84} & \textbf{80.67$\pm$0.52} & \textbf{61.08$\pm$1.65} & \textbf{74.47$\pm$0.67} & \textbf{83.24$\pm$0.75} & 68.09$\pm$1.70 & \textbf{77.81$\pm$0.96} & \textbf{83.33$\pm$0.75} & \textbf{68.10$\pm$2.47} & \textbf{77.97$\pm$1.02} \\
\Xhline{2\arrayrulewidth}
\end{tabular}
\end{table*}

\begin{figure*}[!t]
\centering
\subfigure[]{\includegraphics[width=0.245\textwidth]{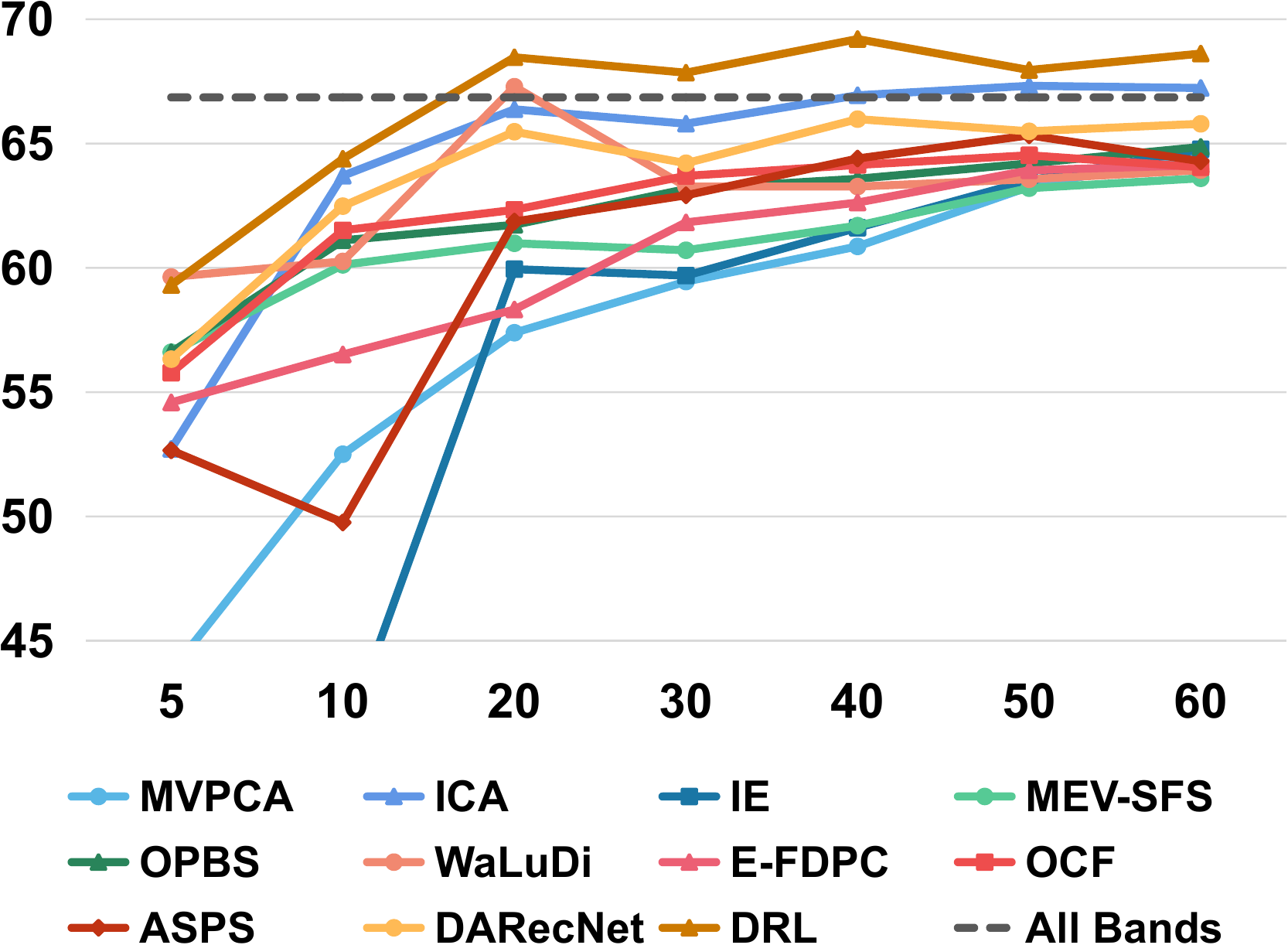}}
\subfigure[]{\includegraphics[width=0.245\textwidth]{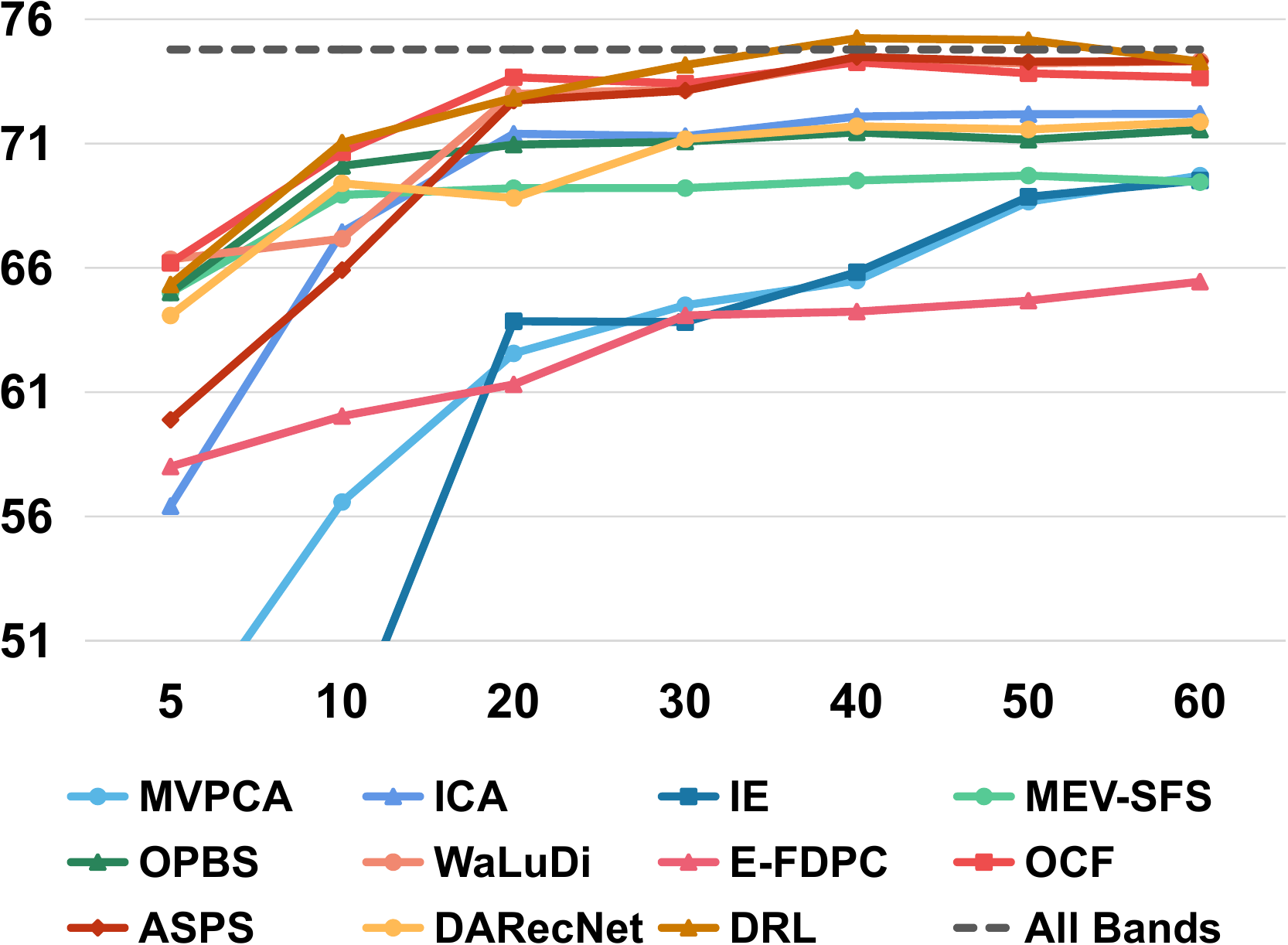}}
\subfigure[]{\includegraphics[width=0.245\textwidth]{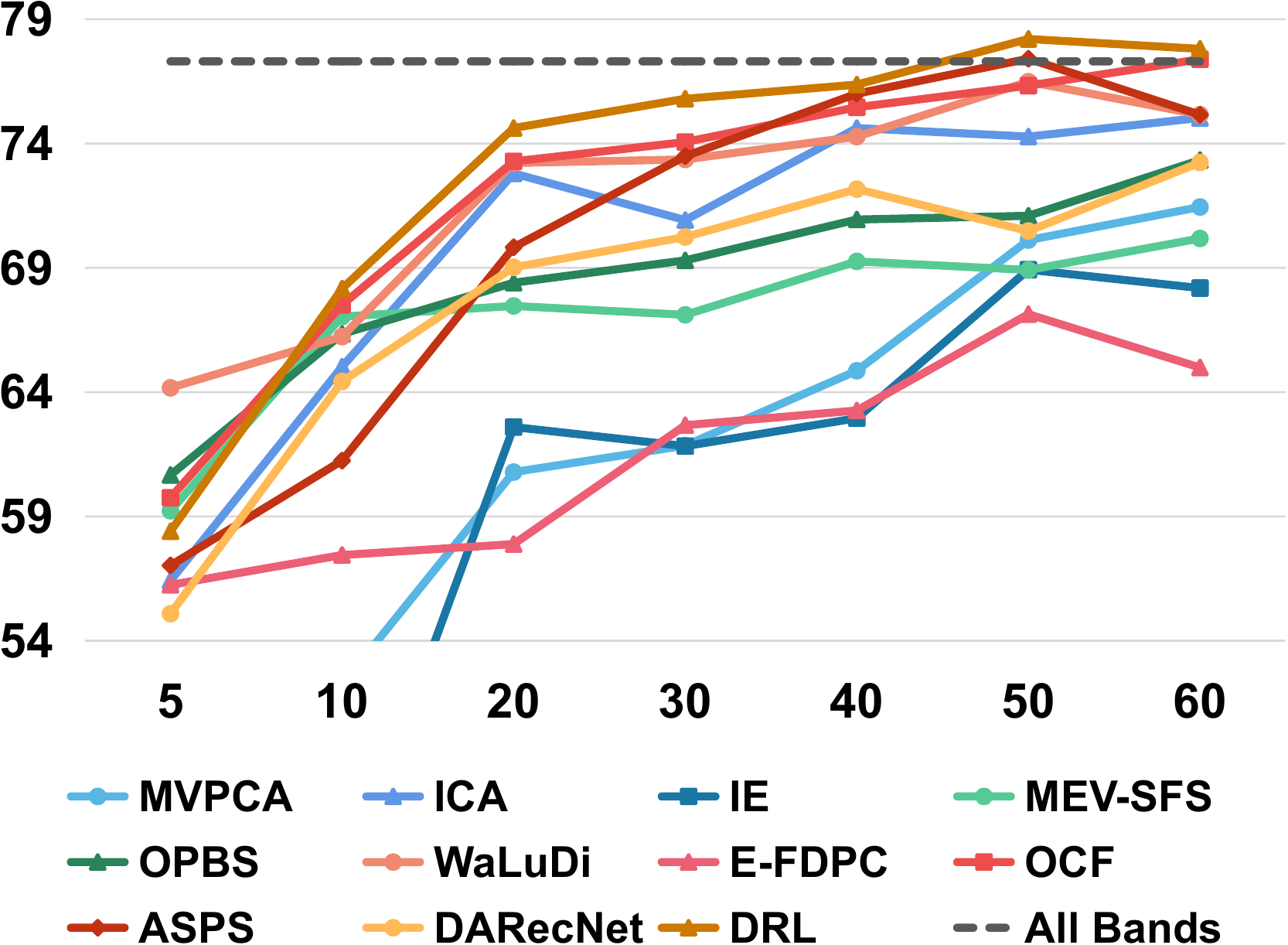}}
\subfigure[]{\includegraphics[width=0.245\textwidth]{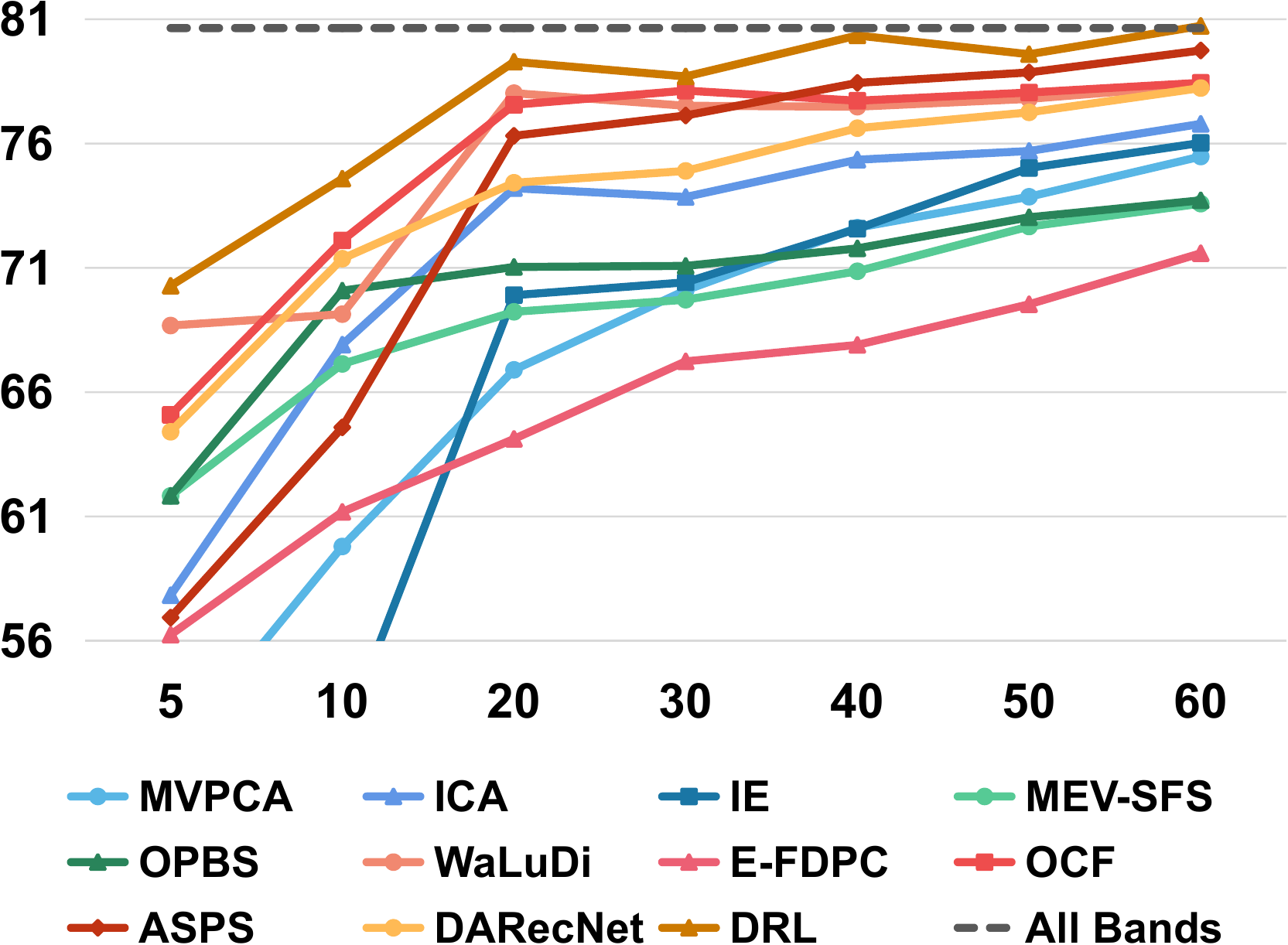}}
\renewcommand{\figurename}{Fig.}
\caption{\label{fig:DiffBands_indian_pines} \RR{}{OA curves of different band selection methods on the Indian Pines data set. The x-axis indicates OA (\%), and the y-axis indicates the number of selected bands. (a) OA by k-NN. (b) OA by RF. (c) OA by MLP. (d) OA by SVM-RBF. All OAs are achieved by averaging 10 individual runs.}}
\end{figure*}

\begin{figure*}[!t]
\centering
\subfigure[]{\label{subfig:1}\includegraphics[width=0.245\textwidth]{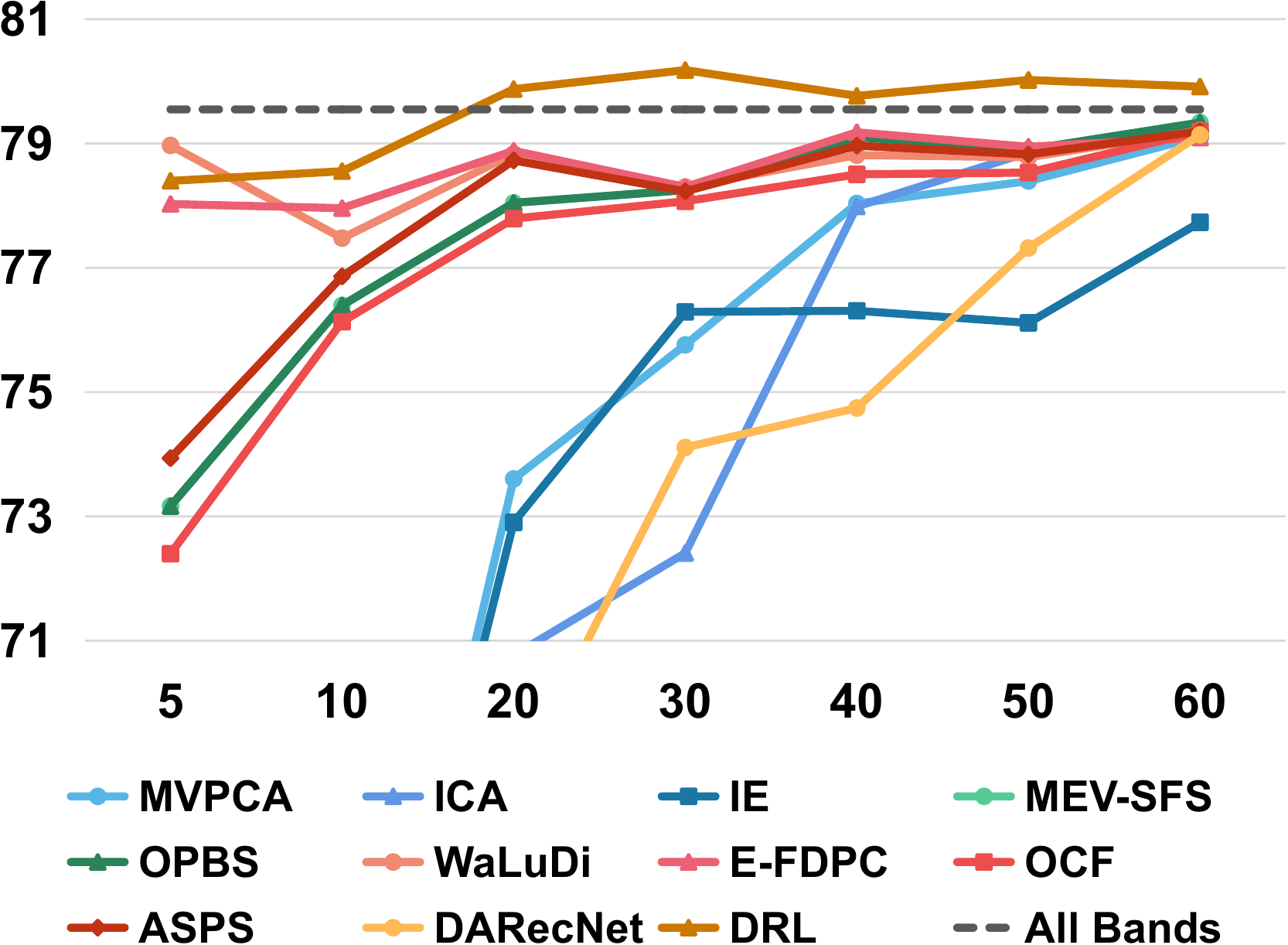}}
\subfigure[]{\label{subfig:2}\includegraphics[width=0.245\textwidth]{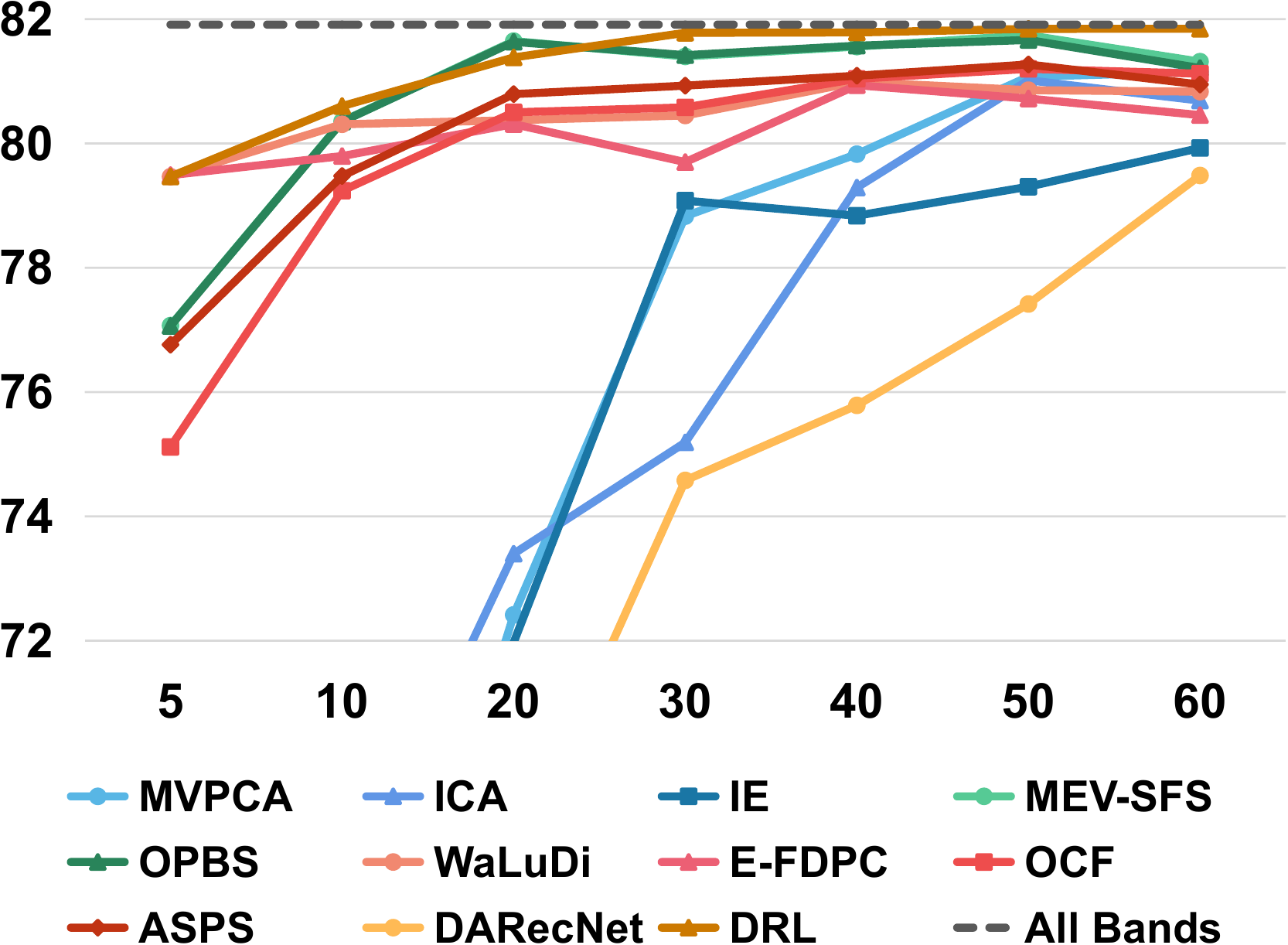}}
\subfigure[]{\label{subfig:25}\includegraphics[width=0.245\textwidth]{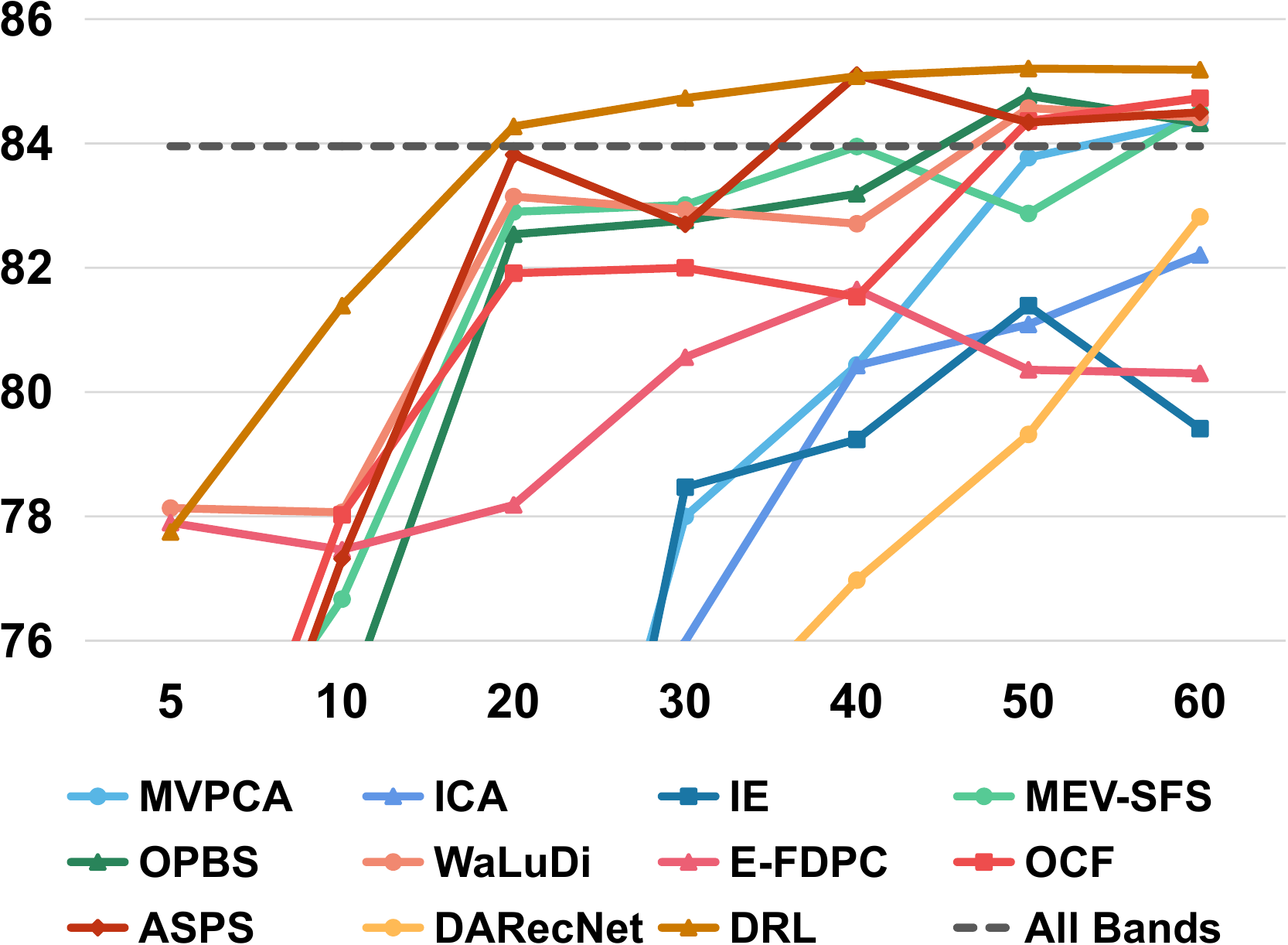}}
\subfigure[]{\label{subfig:3}\includegraphics[width=0.245\textwidth]{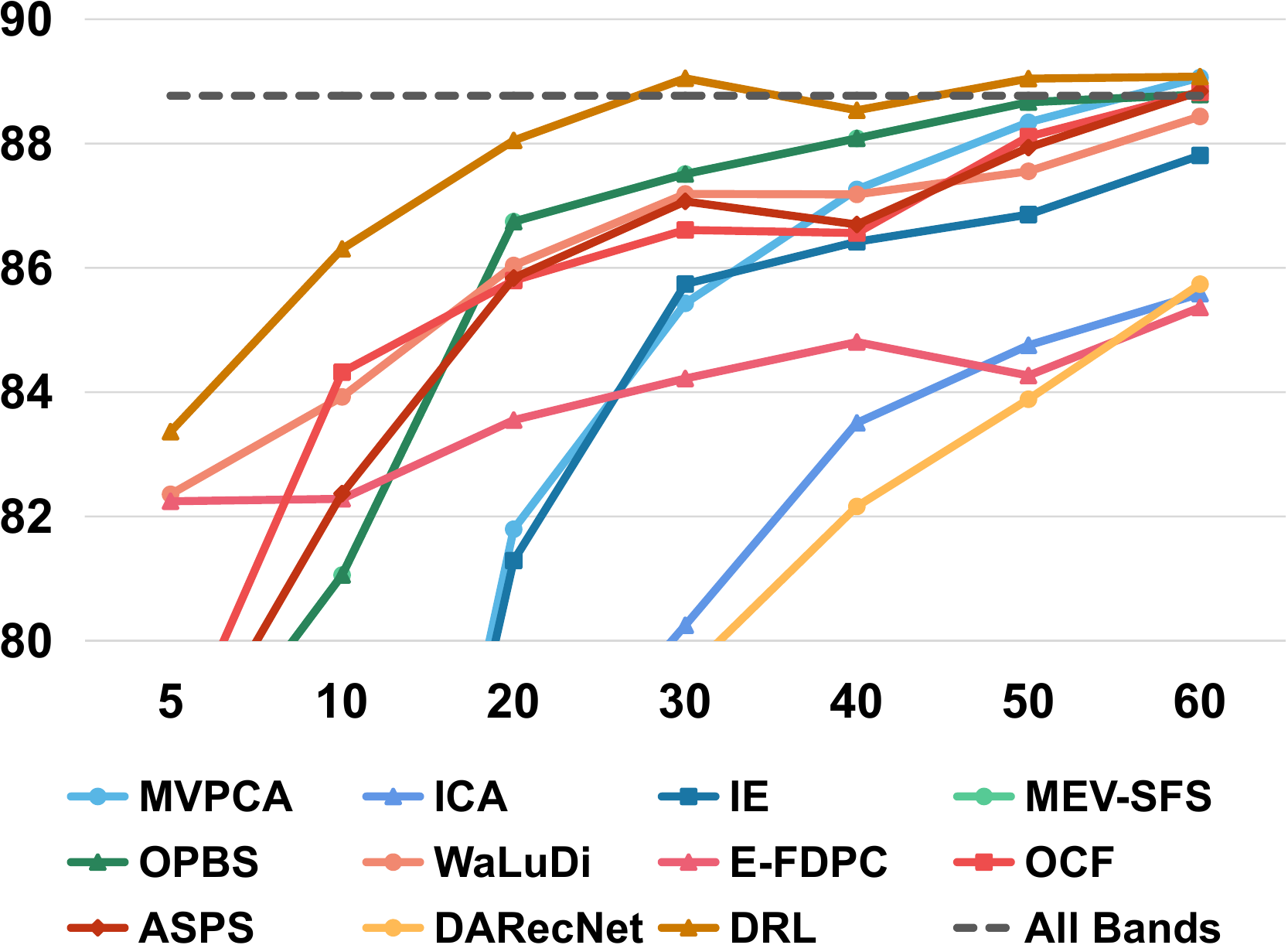}}
\renewcommand{\figurename}{Fig.}
\caption{\label{fig:DiffBands_pavia_uni} \RR{}{OA curves of different band selection methods on the Pavia University data set. The x-axis indicates OA (\%), and the y-axis indicates the number of selected bands. (a) OA by k-NN. (b) OA by RF. (c) OA by MLP. (d) OA by SVM-RBF. All OAs are achieved by averaging 10 individual runs.}}
\end{figure*}

\begin{figure*}[!t]
\centering
\subfigure[]{\label{subfig:4}\includegraphics[width=0.245\textwidth]{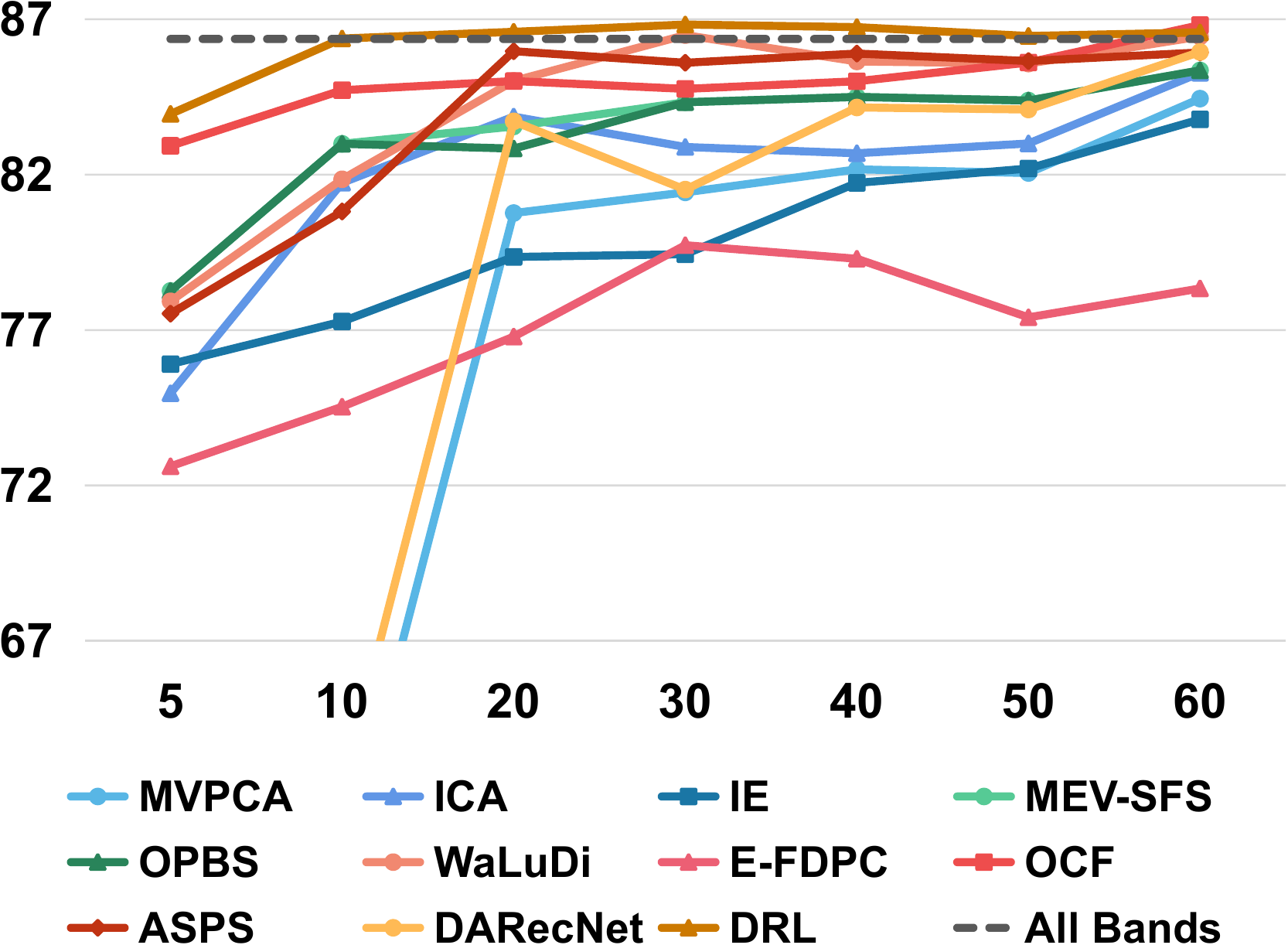}}
\subfigure[]{\label{subfig:5}\includegraphics[width=0.245\textwidth]{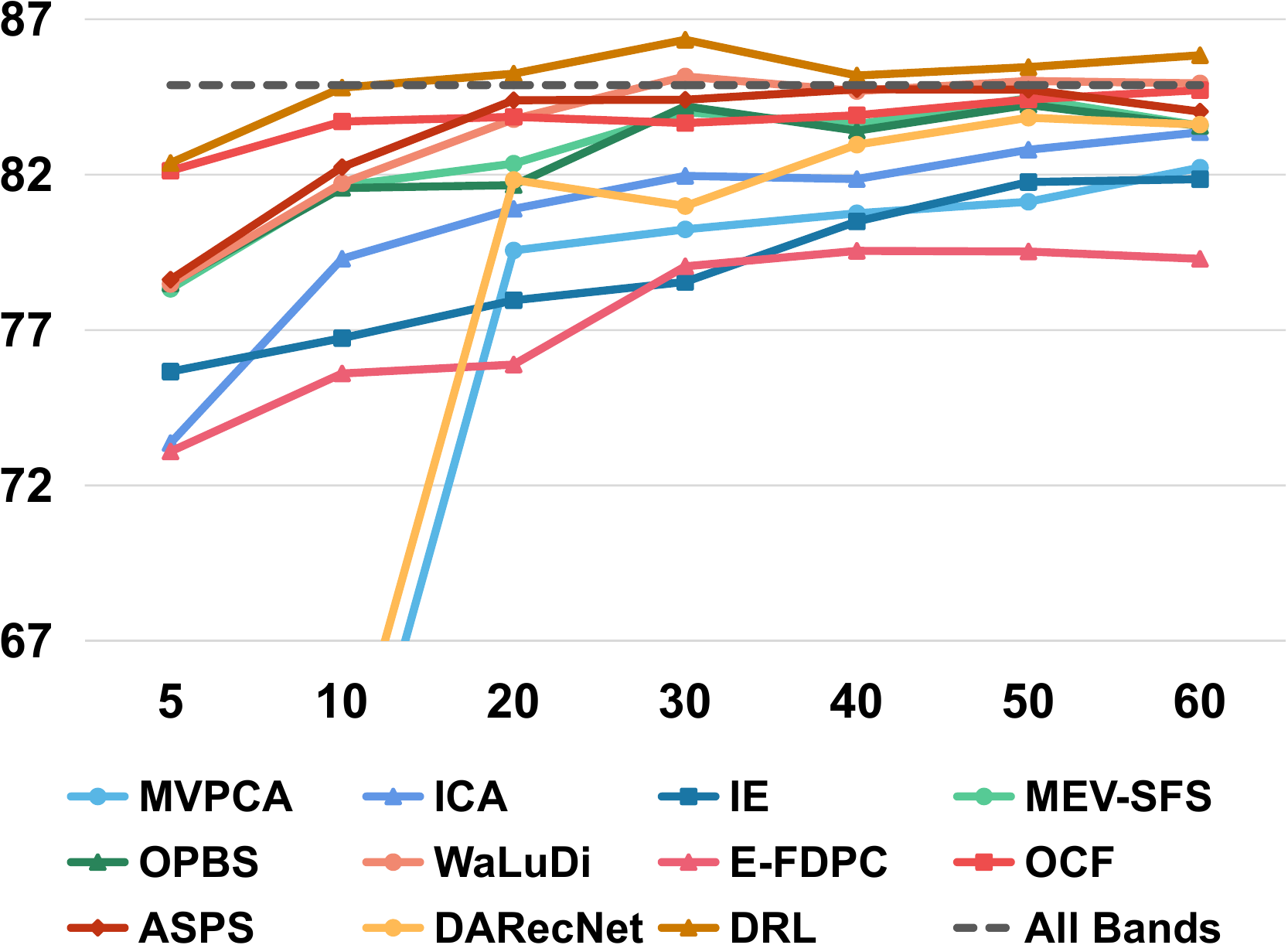}}
\subfigure[]{\includegraphics[width=0.245\textwidth]{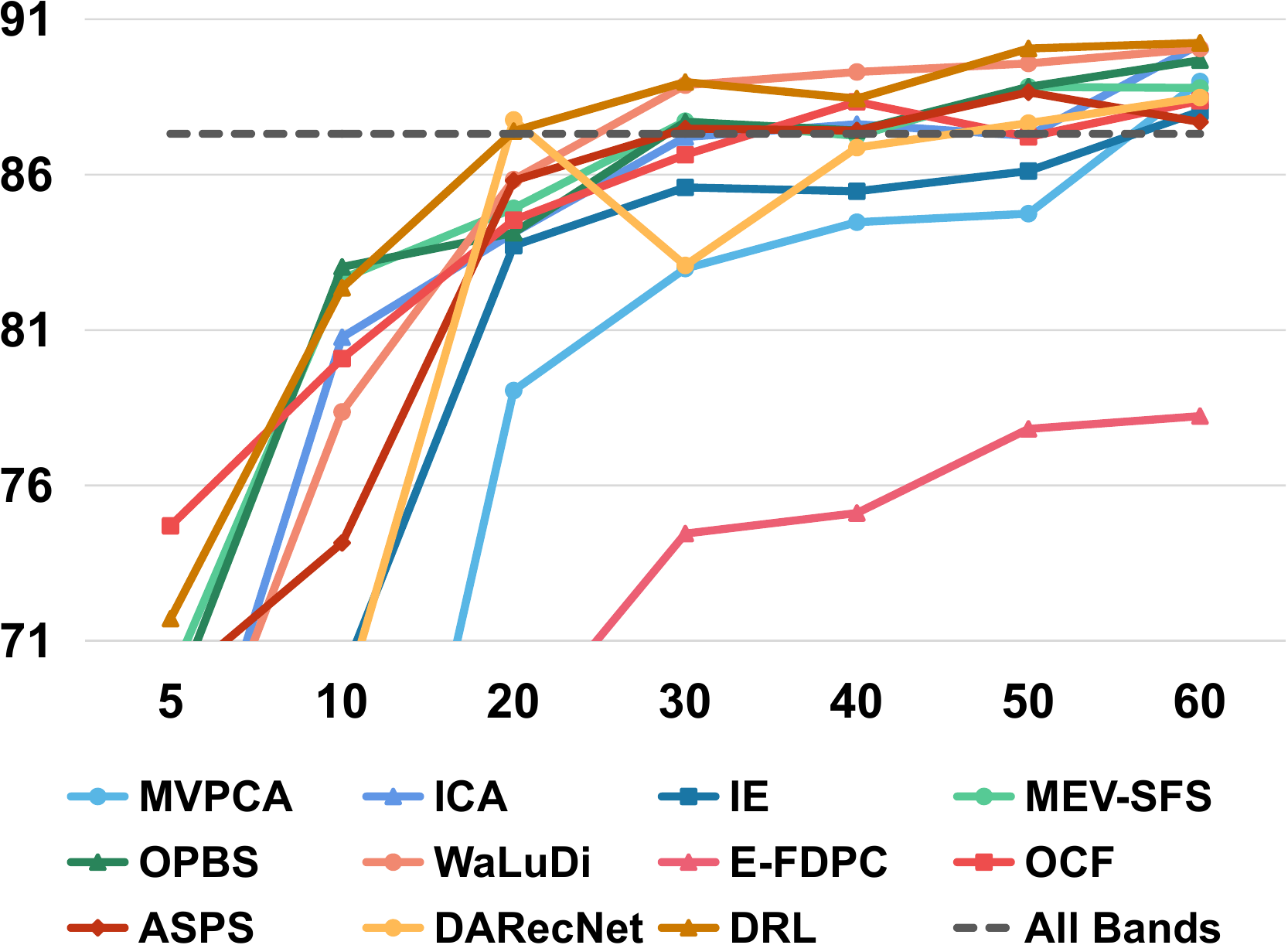}}
\subfigure[]{\label{subfig:6}\includegraphics[width=0.245\textwidth]{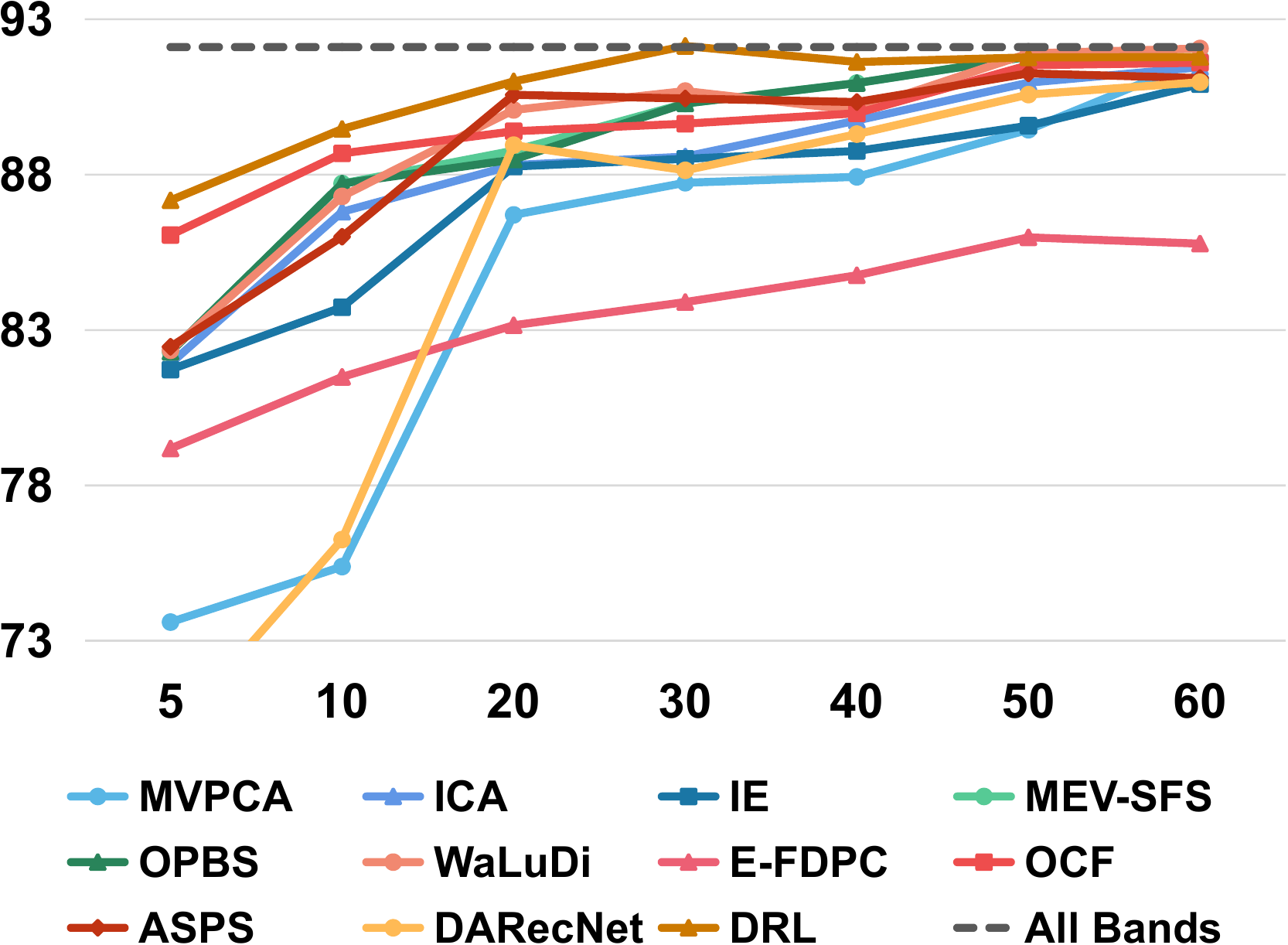}}
\renewcommand{\figurename}{Fig.}
\caption{\label{fig:DiffBands_botswana} \RR{}{OA curves of different band selection methods on the Botswana data set. The x-axis indicates OA (\%), and the y-axis indicates the number of selected bands. (a) OA by k-NN. (b) OA by RF. (c) OA by MLP. (d) OA by SVM-RBF. All OAs are achieved by averaging 10 individual runs.}}
\end{figure*}

\begin{figure*}[!t]
\centering
\subfigure[]{\includegraphics[width=0.245\textwidth]{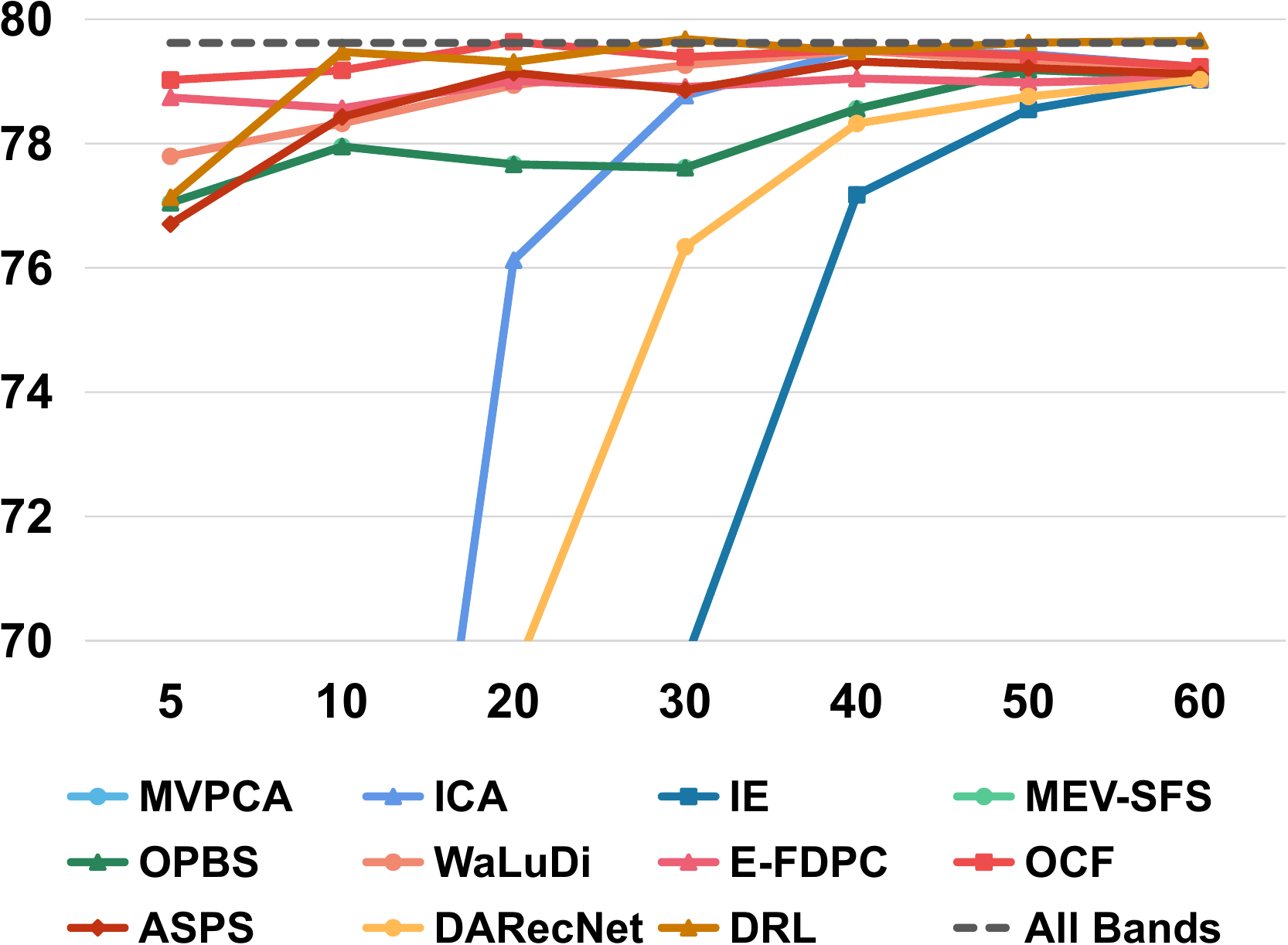}}
\subfigure[]{\includegraphics[width=0.245\textwidth]{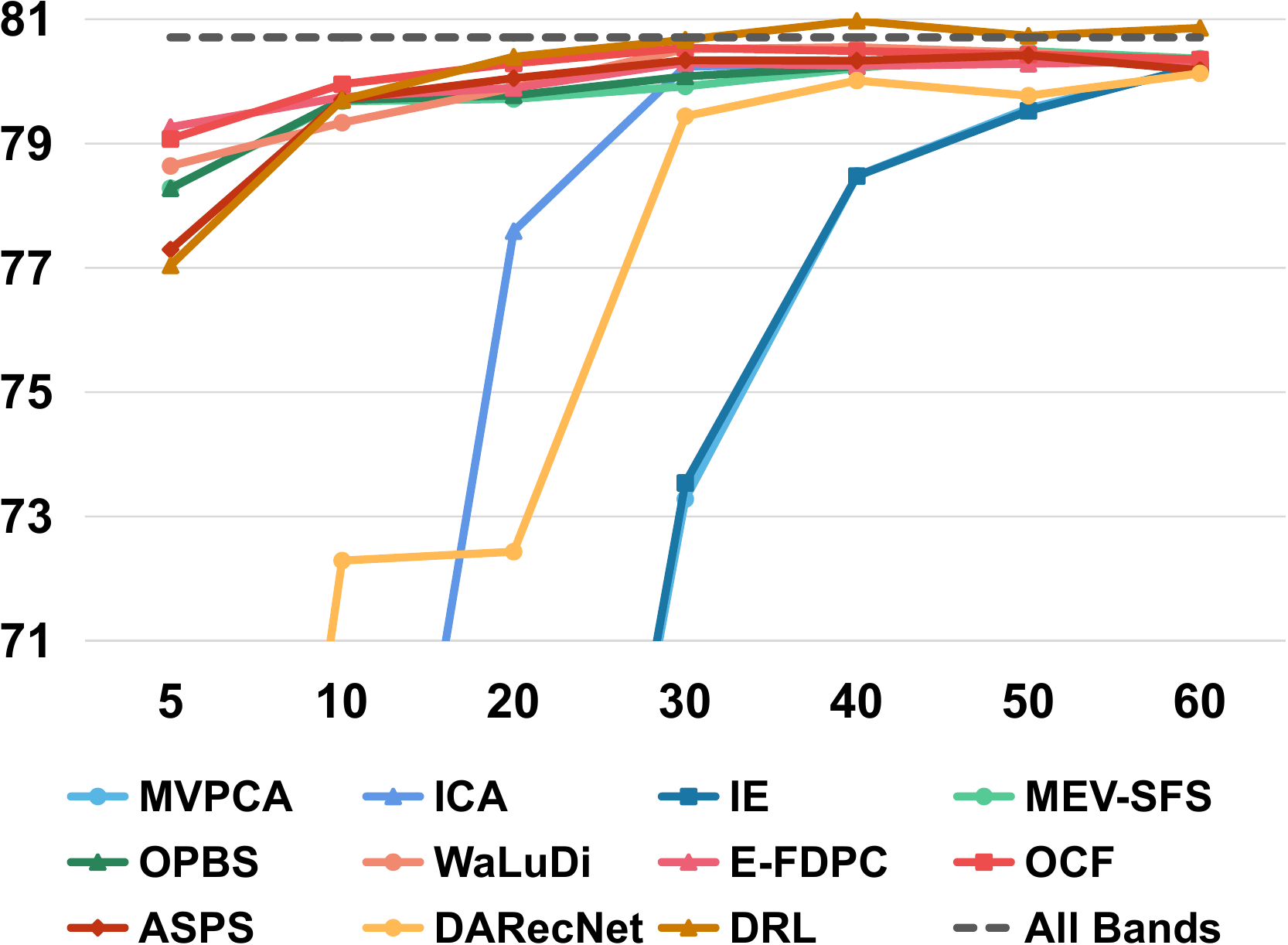}}
\subfigure[]{\includegraphics[width=0.245\textwidth]{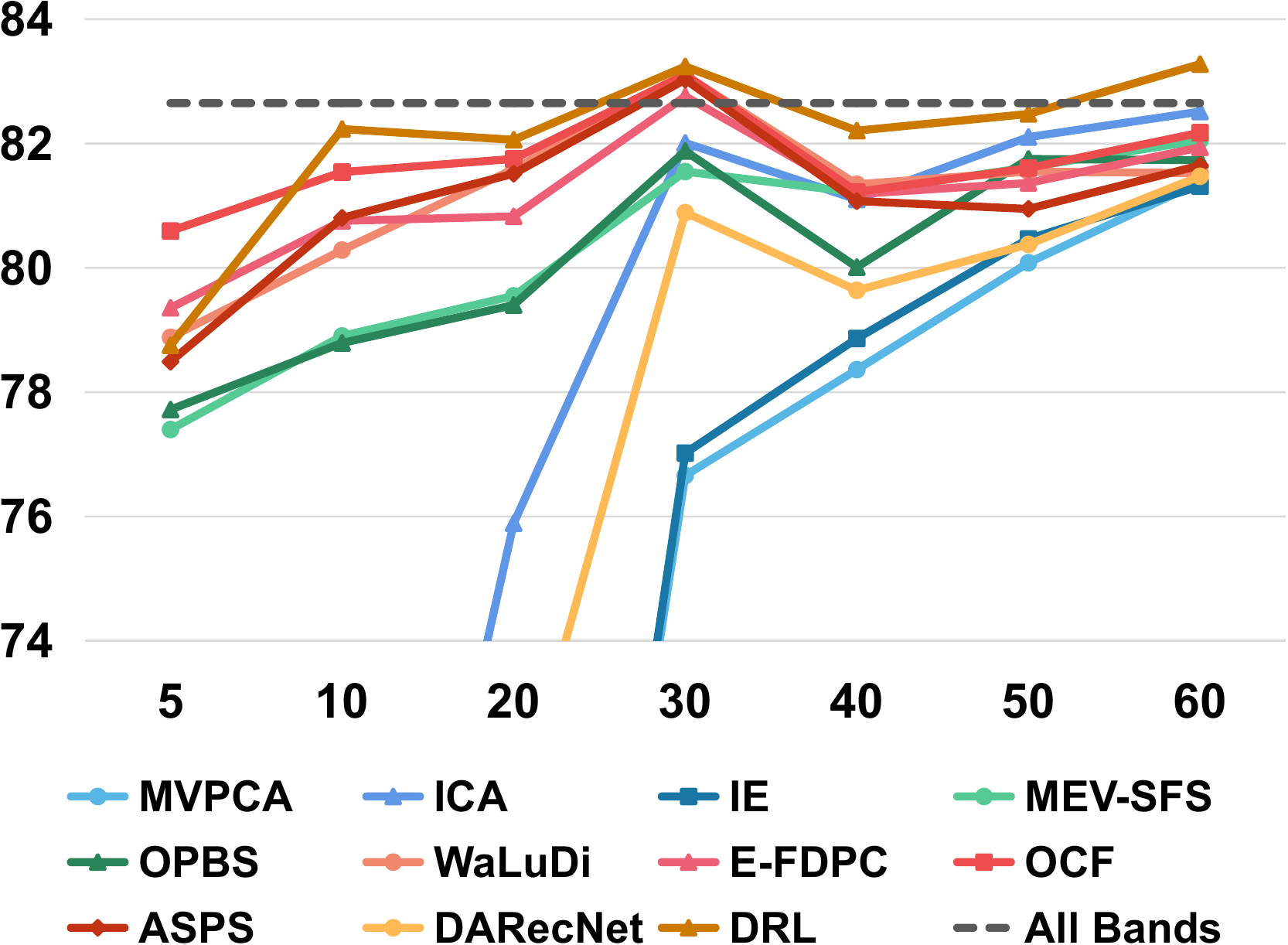}}
\subfigure[]{\includegraphics[width=0.245\textwidth]{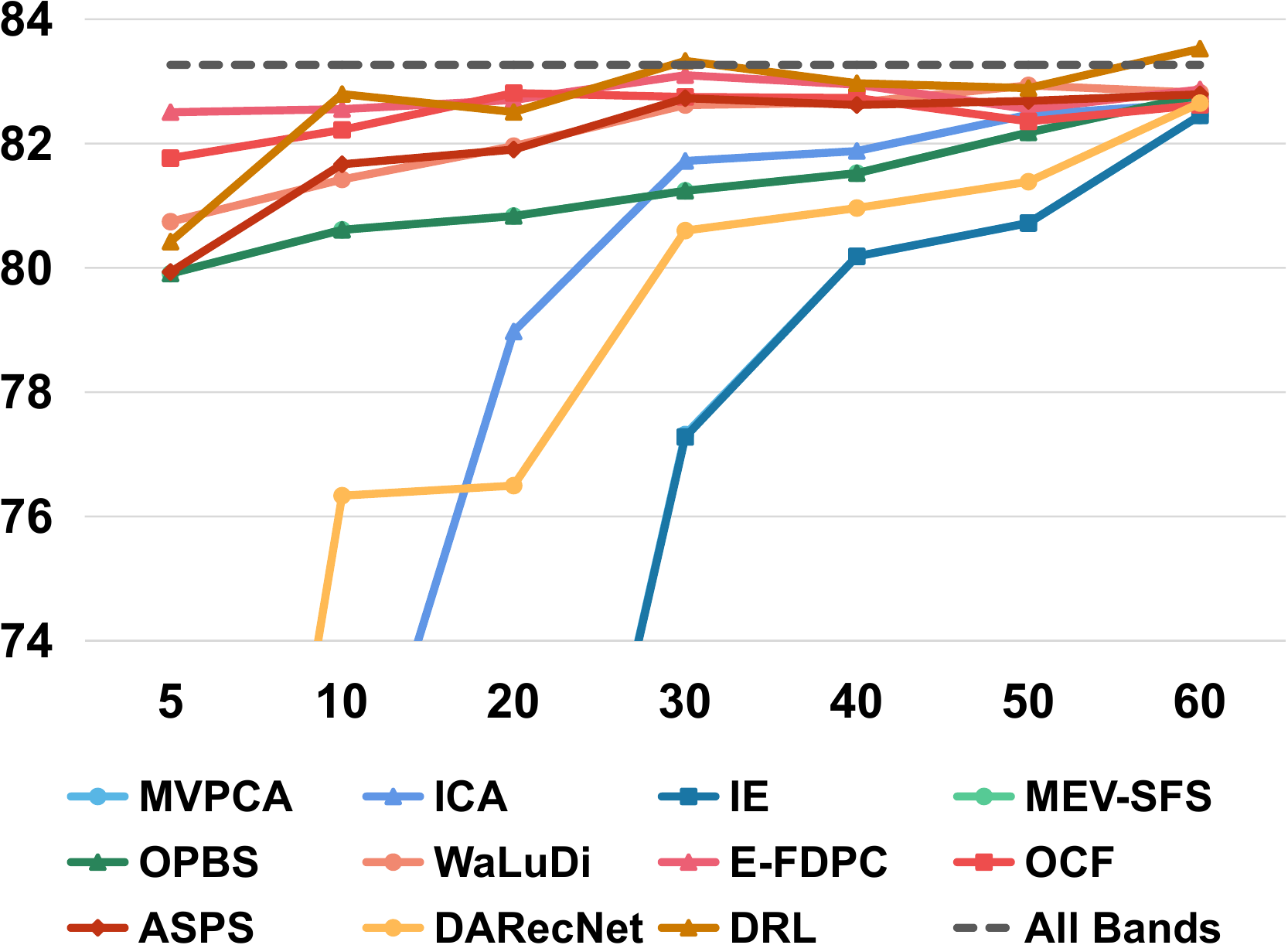}}
\renewcommand{\figurename}{Fig.}
\caption{\label{fig:DiffBands_muufl_gulfport} \RR{}{OA curves of different band selection methods on the MUUFL Gulfport data set. The x-axis indicates OA (\%), and the y-axis indicates the number of selected bands. (a) OA by k-NN. (b) OA by RF. (c) OA by MLP. (d) OA by SVM-RBF. All OAs are achieved by averaging 10 individual runs.}}
\end{figure*}

\subsection{Information Entropy or Correlation: Whose Call Is It in Building The Reward Scheme?}
\label{subsec:exp-c}
Fig.~\ref{fig:IEvsCorr} compares two instantiations of the reward scheme, namely information entropy and correlation coefficient (cf. Section~\ref{subsec:MDP}), on the Pavia University data set. To quantitatively evaluate them, we make use of k-NN to perform classification using spectral bands selected by models using these two schemes. From Fig.~\ref{fig:IEvsCorr}, it can be observed that the former can achieve higher OA, AA, and Kappa coefficient compared to the latter. Moreover, the computation cost of information entropy is lower than that of correlation coefficient. Hence we choose information entropy as the reward scheme in our model for the following experiments.

\begin{figure}[!t]
\centering
\includegraphics[width=\columnwidth]{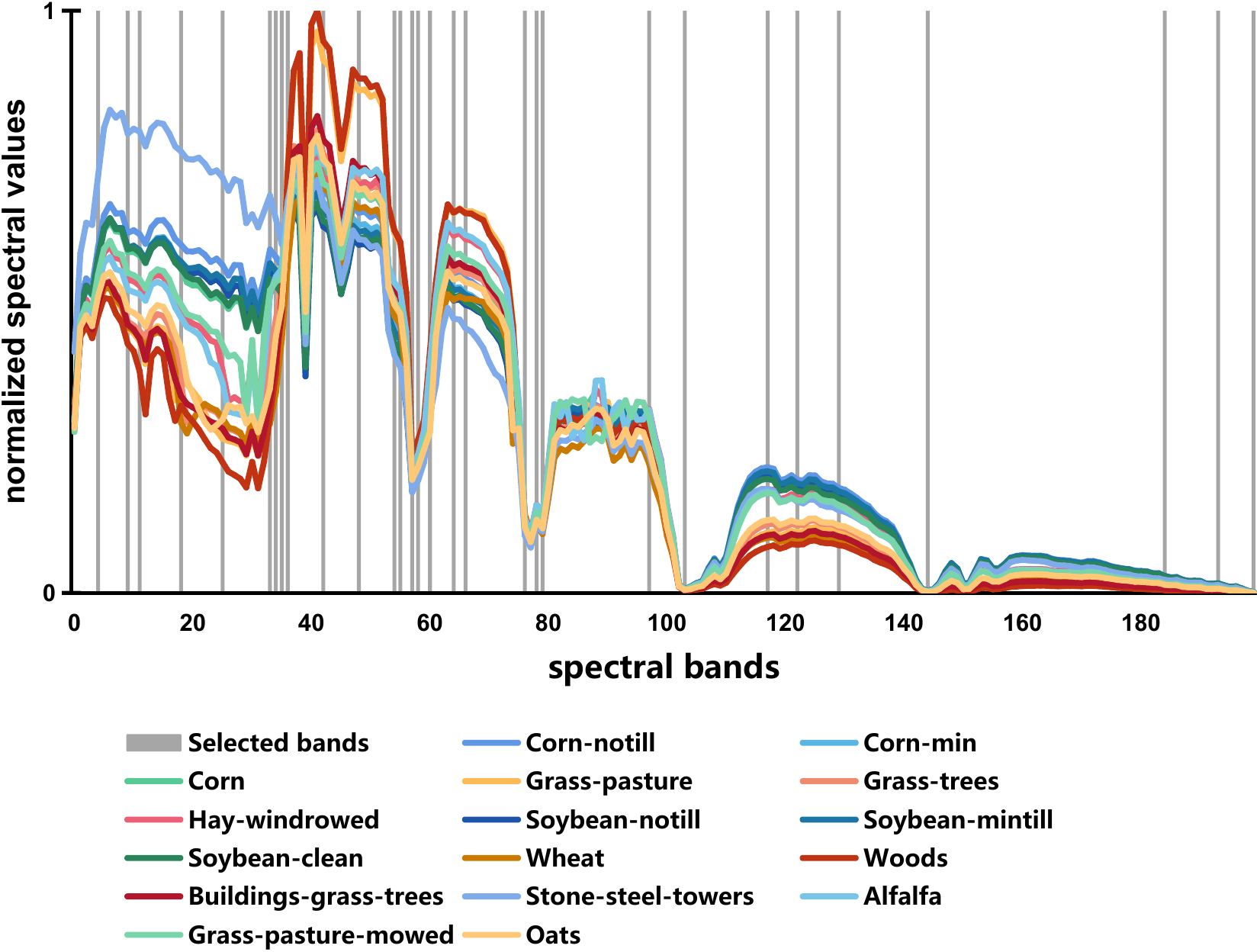}
\renewcommand{\figurename}{Fig.}
\caption{\label{fig:vis_indian_pines} \RR{}{Visualization of bands selected by the proposed method on the Indian Pines data set. We also show the average spectral signature of each class. 30 bands are selected here.}}
\end{figure}

\subsection{Results and Discussion}
In this subsection, we assess the proposed approach by comparing it with several state-of-the-art band selection methods mentioned in Section~\ref{sunsec:exp-b}. For each data set, we plot OA curves showing OA variations w.r.t. the number of chosen bands $K$ in Fig.~\ref{fig:DiffBands_indian_pines}, Fig.~\ref{fig:DiffBands_pavia_uni}, Fig.~\ref{fig:DiffBands_botswana}, and Fig.~\ref{fig:DiffBands_muufl_gulfport}. In our experiments, $K$ varies from 5 to 60. Furthermore, in Table~\ref{tab:indian_pines}, Table~\ref{tab:pavia_uni}, Table~\ref{tab:botswana}, and Table~\ref{tab:muufl_gulfport}, we also report OAs, AAs, and Kappa coefficients of different methods with a fixed $K$ (following the setup in~\cite{WSun19}, it is set to 30).
\par
Fig.~\ref{fig:DiffBands_indian_pines} and Table~\ref{tab:indian_pines} present results on the Indian Pines data set. As can be seen in Fig.~\ref{fig:DiffBands_indian_pines}, the proposed DRL is capable of achieving the highest OA using an SVM classifier with 5 to 60 selected bands. Although the OA of DRL is a little bit lower than that of WaLuDi when a k-NN is employed with 5 bands, our DRL model outperforms other competitors when more spectral bands are chosen. Moreover, we can see that as compared to other band selection models, the proposed approach can also provide gains when using an MLP (the only exception is when $K=5$). With an RF classifier, OCF and WaLuDi outperform DRL when $K=5$ and $20$, but in other cases, the proposed model is able to provide the best results. In Table~\ref{tab:indian_pines}, we take an example of selecting 30 bands for classification and report numerical results. It can be observed that our DRL obtains the best results. Particularly when using a k-NN classifier, our approach can gain an improvement of 2.05\%, 2.13\%, and 2.31\% in OA, AA, and Kappa coefficient, respectively, compared with the second best model. In addition, it is noteworthy that in comparison with original data with all bands, our method can offer almost same or better results at some point, e.g., when over 20 bands are selected for k-NN.
\par
Fig.~\ref{fig:DiffBands_pavia_uni} and Table~\ref{tab:pavia_uni} exhibit classification results for the Pavia University data set. In Fig.~\ref{subfig:1}, Fig.~\ref{subfig:2}, and Fig.~\ref{subfig:25} (with k-NN, RF, and MLP), when only a few bands are selected, e.g., 5, the OA of DRL is lower than that of WaLuDi and/or E-FDPC. But when that number goes beyond 5, the proposed method outperforms other competitors. On the other hand, DRL performs well with an SVM classifier, and its OA exceeds accuracies of all competitors (cf. Fig.~\ref{subfig:3}). For instance, Table~\ref{tab:pavia_uni} shows that as compared to the second best model, MEV-SFS and OPBS, our DRL is able to obtain a gain of 1.54\% and 2.07\% in OA and Kappa coefficient, respectively. Besides, OAs produced by most methods grow when more bands are selected, and our band selection method can achieve higher accuracies than all bands at certain locations, e.g., $K\geq20$ in Fig.~\ref{subfig:2} and $K=30$ and $\geq50$ in Fig.~\ref{subfig:3}.

\begin{figure}[!t]
\centering
\includegraphics[width=\columnwidth]{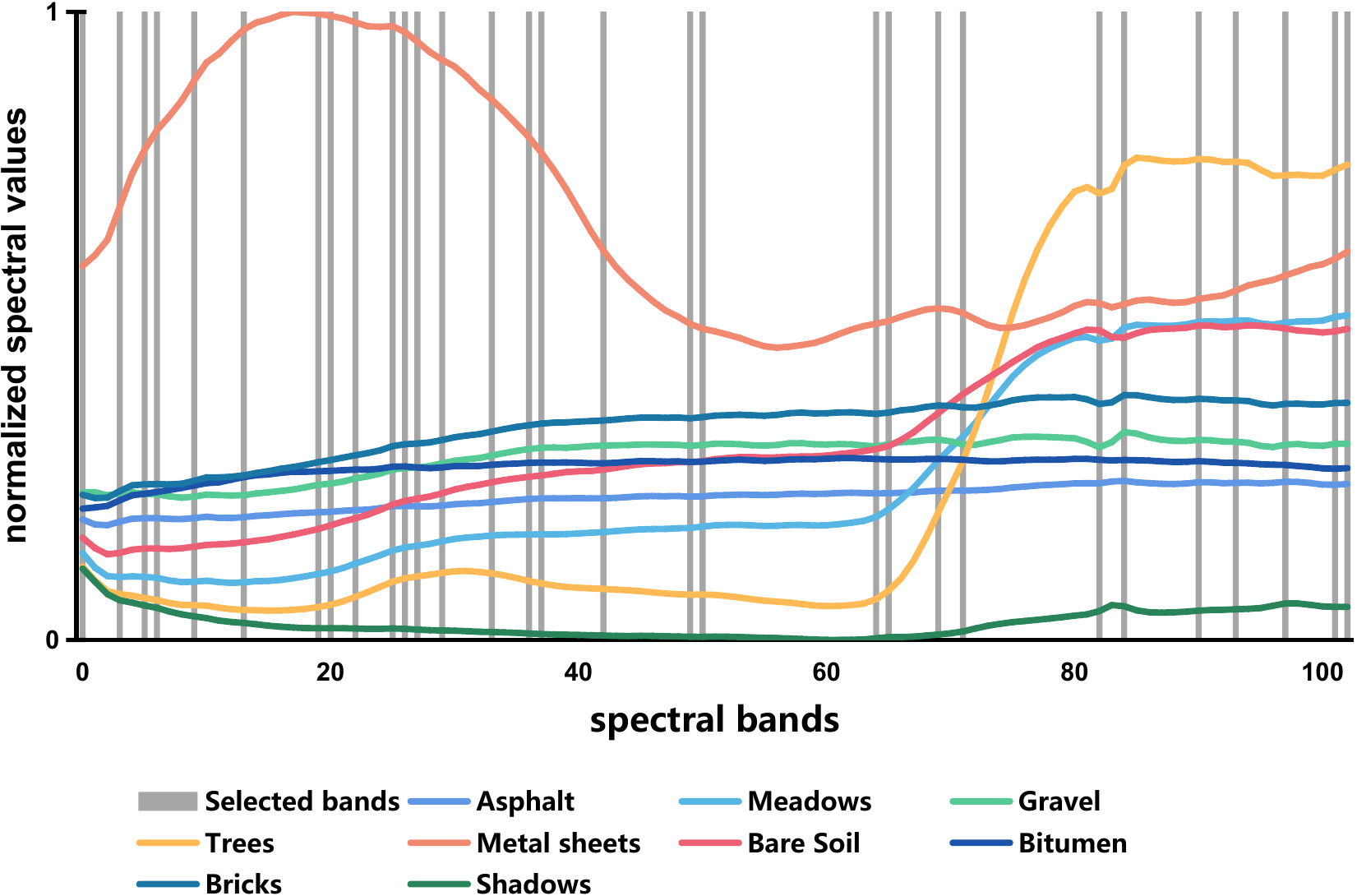}
\renewcommand{\figurename}{Fig.}
\caption{\label{fig:vis_pavia_uni} \RR{}{Visualization of bands selected by the proposed method on the Pavia University data set. We also show the average spectral signature of each class. 30 bands are selected here.}}
\end{figure}

Classification results on the Botswana data set are shown in Fig.~\ref{fig:DiffBands_botswana} and Table~\ref{tab:botswana}. As shown in Fig.~\ref{fig:DiffBands_botswana}, the proposed DRL model delivers the best and most stable results with k-NN, RF, and SVM, except for $K=60$ in Fig.~\ref{subfig:4} and Fig.~\ref{subfig:6}. For MLP, DRL performs best when $K=30$, $50$, and $60$, and in other cases, it achieves the second best results. Besides, we notice that the performance of DRL and some competitors is very similar when selecting more than 50 spectral bands. But overall we can see significant gains on this data set.
\par
In Fig.~\ref{fig:DiffBands_muufl_gulfport} and Table~\ref{tab:muufl_gulfport}, we report results on the MUUFL Gulfport data set. It can be seen that several band selection models, e.g., WaLuDi, E-FDPC, OCF, ASPS, and DRL, behave very similarly. This may be because as compared to the other three data sets, the MUUFL Gulfport data set has only 64 bands.
\par
\RR{}{Overall, from the tables, we can see that among all band selection models, the ranking-based methods perform relatively poorly, while the clustering-based approaches tend to achieve good results. The searching-based models, i.e., MEV-SFS and OPBS, can deliver good selected bands on some data sets like the Pavia University scene, but it is noteworthy that they are not robust against different data sets, for example, their performance on the Indian Pines scene is not satisfactory. By contrast, our method shows superior performance.} \RR{}{This may be due to the fact that our approach is a data- and objective-driven learning-based model. Compared to other heuristic algorithms, it is able to explore more possible band subsets during the training phase.}
\par
\begin{figure}[!t]
\centering
\includegraphics[width=\columnwidth]{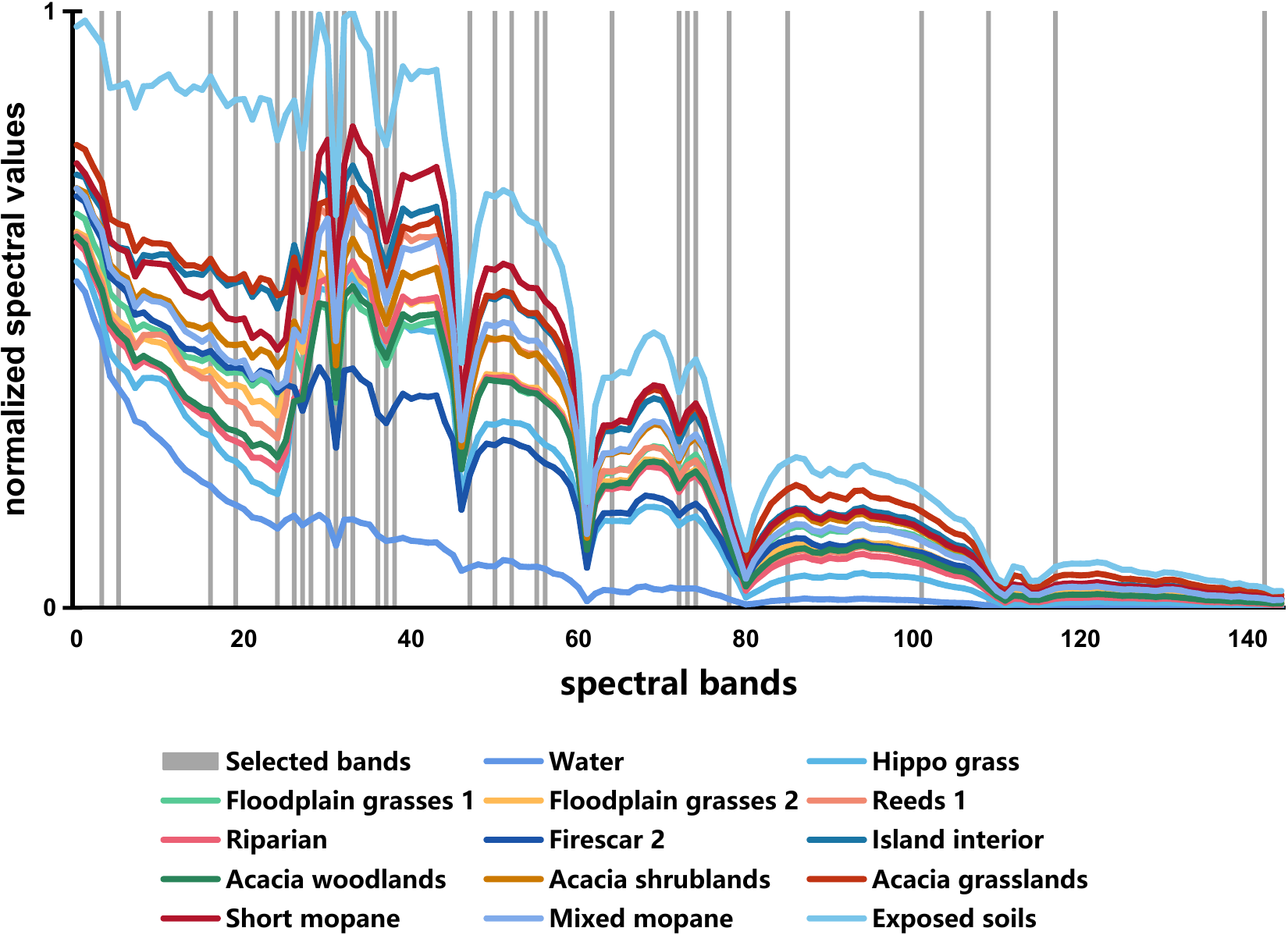}
\renewcommand{\figurename}{Fig.}
\caption{\label{fig:vis_botswana} \RR{}{Visualization of bands selected by the proposed method on the Botswana data set. We also show the average spectral signature of each class. 30 bands are selected here.}}
\end{figure}
\begin{figure}[!t]
\centering
\includegraphics[width=\columnwidth]{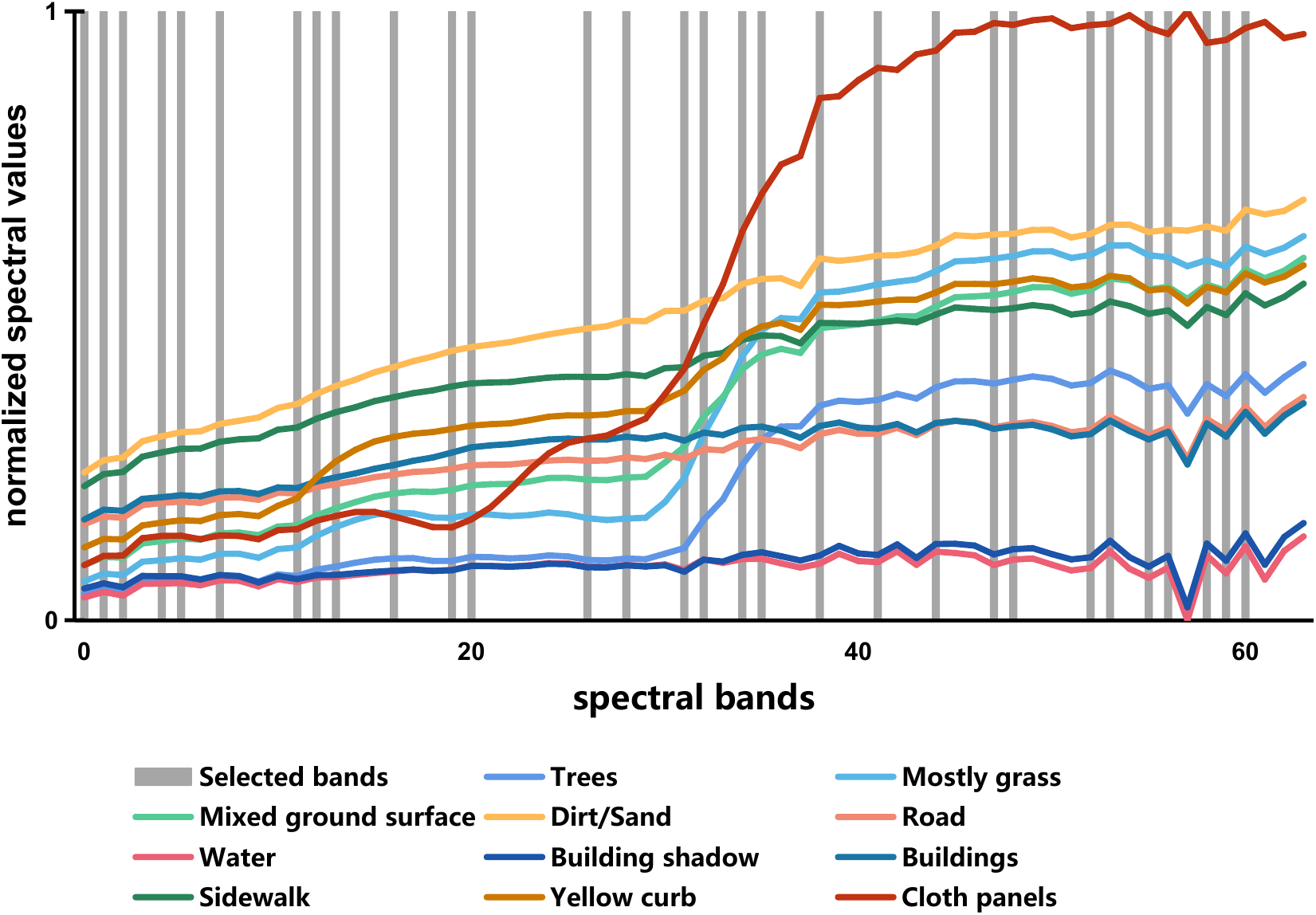}
\renewcommand{\figurename}{Fig.}
\caption{\label{fig:vis_muufl_gulfport} \RR{}{Visualization of bands selected by the proposed method on the MUUFL Gulfport data set. We also show the average spectral signature of each class. 30 bands are selected here.}}
\end{figure}
\begin{figure}[!t]
\centering
\includegraphics[width=0.9\columnwidth]{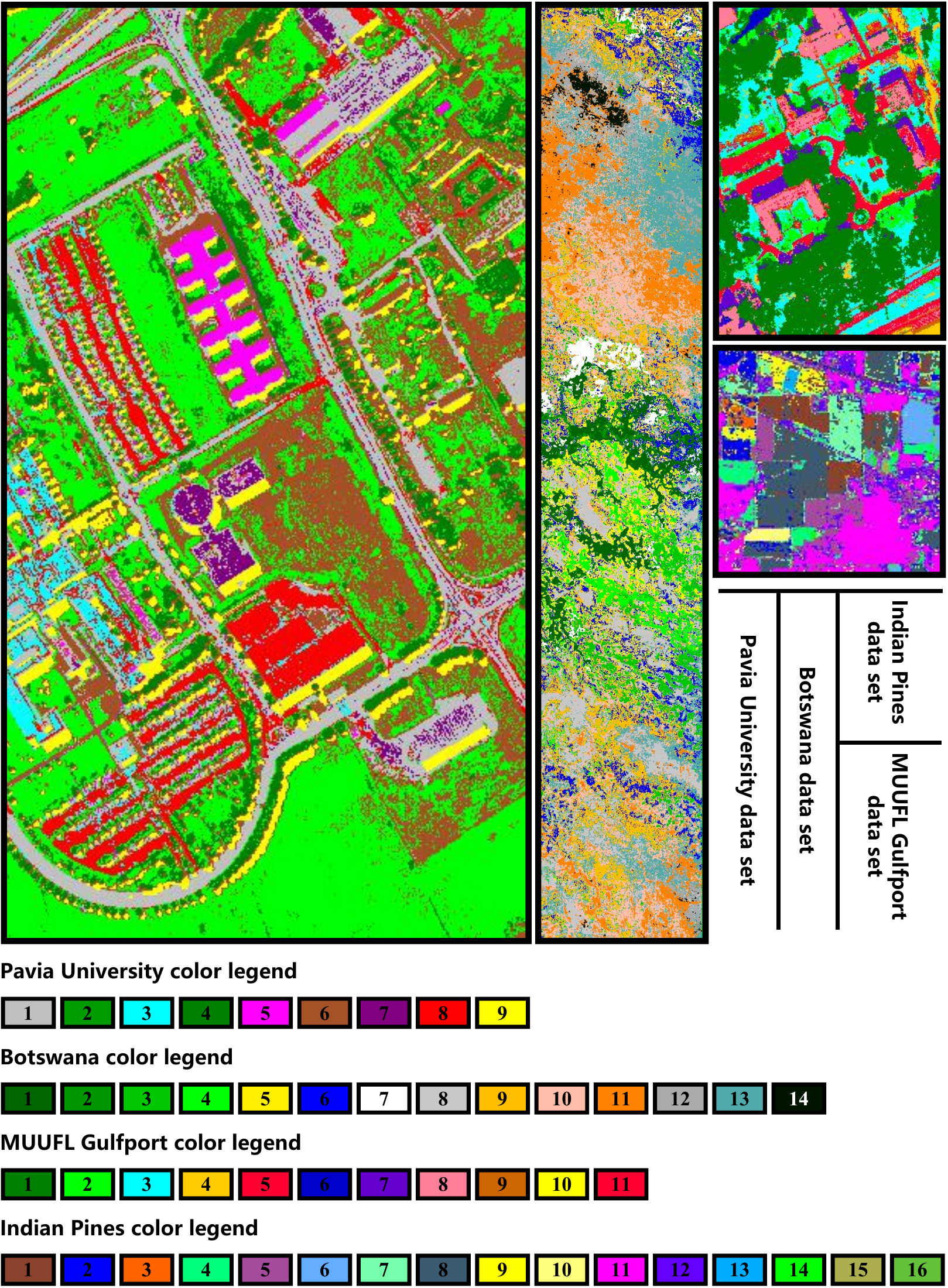}
\renewcommand{\figurename}{Fig.}
\caption{\label{fig:classification_maps} \RR{}{Classification maps using 30 bands selected by the proposed DRL model and an SVM-RBF classifier on the four data sets.}}
\end{figure}

In addition, we visualize bands selected by the proposed method on both four data sets in Fig.~\ref{fig:vis_indian_pines}, Fig.~\ref{fig:vis_pavia_uni}, Fig.~\ref{fig:vis_botswana}, and Fig.~\ref{fig:vis_muufl_gulfport}. \RR{}{From these figures, we see that DRL tends to select spectral bands with high information entropy. This is in line with our presumption and existing studies in hyperspectral band selection, in which information entropy is an important measurement.} Classification maps using 30 bands selected by the proposed DRL model and an SVM-RBF classifier on the four data sets are shown in Fig.~\ref{fig:classification_maps}. \RR{}{Basically, these maps present satisfactory classification results, although we see some salt-and-pepper noise which are inevitable in spectral classification.}

\subsection{Stability against Classifiers and Robustness against Data Sets}
From experimental results, we observe that some competitors have unstable behaviors with different classifiers. For example, ASPS works quite well on the Indian Pines data set when RF, MLP, and SVM are employed but a little bit poor when using a k-NN. Similarly, E-FDPC can provide decent results on the Pavia University data set with k-NN, while with an MLP or SVM classifier, it performs rather poorly as compared to other band selection algorithms. This is probably because there exist noisy bands in selected bands, which leads to an unsatisfactory performance on noise-sensitive classifiers. In contrast to most competitors, the proposed DRL model is more stable against classifiers.
\par
Furthermore, we also notice that the robustness of several competitors against different data sets is not satisfactory. For example, when 30 bands are selected and making use of an SVM classifier, OCF is capable of achieving the second highest OA and Kappa coefficient on the Indian Pines data set (cf. Table~\ref{tab:indian_pines}) but shows a lackluster performance on the Pavia University and Botswana data sets (see Table~\ref{tab:pavia_uni} and Table~\ref{tab:botswana}). This may be because choosing an optimal combination of spectral bands is a non-trivial task, and locally optimal solutions are not easy to always avoid. In this aspect, the proposed method is more robust against data sets.

\subsection{\RR{}{Limitations}}
\RR{}{Further, we would like to discuss limitations of the proposed method. Firstly, as to computational time, compared to other heuristic band selection methods, the proposed model needs more time, as it is a learning-based algorithm and takes some time to explore an effective band-selection policy during the training phase. Taking the Indian Pines data set and 30 selected bands as an example, most heuristic band selection approaches take a few seconds to several tens of seconds~\cite{sun2020fast}, and the proposed model needs around 350 seconds. But we note that DARecNet~\cite{Roy2020}, a CNN-based unsupervised band selection model, takes about 9000 seconds under recommended settings. Overall, the computational time of our model is acceptable. Secondly, since the objective function of our unsupervised DRL is structured such that the learning is aiming to maximize the reward rather than classification accuracy, we cannot intuitively assess the quality of the model in terms of classification accuracy during the training phase, which may lead to unstable model training and the inconvenience of monitoring model training.}

\section{Conclusion \RR{}{and Outlook}}
\label{sec:conc}
This paper proposes a deep reinforcement learning model for unsupervised hyperspectral band selection. In the training phase, the goal of the deep reinforcement learning agent (i.e., Q-network) is to learn a band-selection policy that guides the sequential decision-making process of this agent. The policy is a function specifying the band to be chosen given the current state. Note that the training process does not need any labeled data. In the test phase, the agent acts sequentially according to the learned policy. We conduct extensive experiments, and results show the effectiveness of our approach. Moreover, two instantiations of the reward scheme in Section~\ref{subsec:MDP} are quantitatively compared, and we believe that more alternatives are possible and may improve results.
\par
In the future, several studies intend to be carried out. For example, combining deep reinforcement learning and some heuristic band selection frameworks (e.g., the clustering-based method) is likely to offer better band selection solutions. \RR{}{Considering that different classes may have different optimal band subsets (with a variable number of bands), how to determine the best band combination for each category is an interesting but challenging problem. A supervised deep reinforcement learning model may be able to provide insights.} Moreover, we believe that deep reinforcement learning can be applied to more remote sensing applications, such as multitemporal data analysis, visual reasoning in airborne or space-borne images, and other combinatorial optimization tasks in remote sensing.

\section*{Acknowledgement}
The authors would like to thank F. Zhang for helpful discussions on this paper.

\ifCLASSOPTIONcaptionsoff
  \newpage
\fi

\bibliographystyle{IEEEtran}
\bibliography{IEEEfull,reference}

\end{document}